\documentclass{ptephy_v1}
\usepackage{graphicx}      
\usepackage{pstricks}                                                                   
\usepackage{pstricks-add}
\newcommand{\Fh}[2]{\,{}_#1F_#2}
\newcommand{\Fs}[3]
{\!\!\left[\begin{array}{c}#1\,;\\#2\,;
\end{array}#3\right]}
\newcommand{\Fz}[3]{\Fs{#1}{#2}{#3}}
\begin{document}
\title{One-loop three-point Feynman integrals 
with Appell $F_1$ hypergeometric functions}
\author{Khiem Hong Phan and 
Dzung Tri Tran}
\affil{VNUHCM-University of 
Science,~$227$ Nguyen Van Cu, Dist.~$5$, Ho
Chi Minh City, Vietnam
\email{phkhiem@hcmus.edu.vn}}
\begin{abstract}%
New analytic formulas for one-loop three-point 
Feynman integrals in general space-time dimension
($d$) are presented in this paper. The calculations 
are performed at general configurations for internal 
masses and external momenta. The analytic results are 
expressed in terms of hypergeometric series 
$_2F_1$, $_3F_2$ for special cases and Appell $F_1$
for general cases. Furthermore, 
we cross-check our analytic results with other 
references which have carried out the integrals 
in several special cases.
\end{abstract}
\subjectindex{B87}
\maketitle
\section{Introduction}
\noindent
Future experimental programs at 
the High-Luminosity Large Hadron 
Collider (HL-LHC)~\cite{ATLAS:2013hta,
CMS:2013xfa} and the International 
Linear Collider (ILC)~\cite{Baer:2013cma} 
aim to measure precisely the properties 
of Higgs boson, top quark and vector 
bosons for discovering the nature of 
the Higgs sector as well as for finding 
the effects of physics beyond the standard 
model. In order to match the high-precision 
of experimental data in the near future, 
theoretical predictions including 
high-order corrections are required. 
In this framework, detailed evaluations 
of one-loop multi-leg and higher-loop at 
general scale and mass assignments 
are necessary. 

One-loop Feynman integrals in general 
space-time dimension play a crucial 
role for several reasons. Within the 
general framework for computing
two-loop or higher-loop corrections, 
higher-terms in 
the $\varepsilon$-expansion 
(with $\varepsilon =2-d/2$)
from one-loop integrals are necessary for 
building blocks. Moreover, one-loop 
integrals in $d>4$ may be taken into 
account in a reduction for tensor one-loop
\cite{Davydychev:1991va, Fleischer:2010sq}, 
two-loop and higher-loop integrals~\cite{IBP}. 
There have been available many calculations 
for scalar one-loop functions in general 
dimension $d$~\cite{Boos:1990rg,
Davydychev:1990cq,Davydychev:1997wa,
Anastasiou:1999ui,Suzuki:2003jn,Abreu:2015zaa,
Phan:2017xsj}.
However, not all of the calculations 
cover general $\varepsilon$-expansion at general 
scale and internal mass assignments.
Furthermore, a recurrence relation
in $d$ for Feynman loop integrals has been proposed
\cite{Tarasov:1996br} and solved for scalar one-loop 
integrals which have been expressed in terms of 
generalized hypergeometric series 
\cite{Fleischer:2003rm}. 
However, the general solutions for arbitrary 
kinematics have not been found, 
as pointed out in \cite{Bluemlein:2015sia}.
More recently, scalar 1-loop Feynman 
integrals as meromorphic functions in general 
space-time dimension, for arbitrary kinematics 
has been presented 
in~\cite{Bluemlein:2017rbi,Phan:2018cnz}.
In the present paper, new 
analytic formulas for one-loop three-point 
Feynman integrals in general space-time 
dimension are reported by following 
an alternative approach. 
The analytic results are expressed 
in terms of $_2F_1$, $_3F_2$ and
Appell $F_1$ hypergeometric 
functions. The evaluations are performed 
in general configuration of internal masses 
and external momenta. Last but not least, 
our results are cross-checked to other papers 
which have been available in several 
special cases.

The layout of the paper is as follows: In section $2$, 
we present in detail the method for evaluating scalar 
one-loop three-point functions. In this section, 
we first introduce notations used for this work. 
We next consider the case of one-loop triangle diagram 
with two light-like external momenta and 
generalize the calculations for general case. 
Finally, tensor one-loop three-point integrals are 
discussed. Conclusions and outlooks are devoted in 
section $3$. Several useful formulas applied
in this calculation can be found in the appendixes.
 \section{Analytic formulas}
Detailed evaluations for one-loop
three-point integrals 
are presented in this section.  
\subsection{Definitions}
\begin{figure}[ht]
\begin{center}
\begin{pspicture}(-3, -3)(3.5, 3)
\psset{linewidth=0.1pt}
\psset{unit = 0.6}
\psline(-2.5,-2)(0,2)
\psline{->}(-2.5, -2)(-1.25, 0)
\psline(0,2)(2.5,-2)
\psline{->}(0, 2)(1.25, 0)
\psline(-2.5,-2)(2.5,-2)
\psline{->}(2.5, -2)(0, -2)
\psline(-2.5,-2)(-5,-4)
\psline{->}(-5, -4)(-3.8, -3)
\psline(2.5,-2)(5,-4)
\psline{->}(5,-4)(3.8,-3)
\psline(0,2)(0,5)
\psline{->}(0,5)(0,3)
\rput(0.8, 3.5){$p_2^2$}
\rput(-4.2,-2.5){$ p_1^2$}
\rput(4.2, -2.5){$ p_3^2$}
\rput(0, -2.8){$m_3^2$}
\rput(-2.5,0){$m_1^2$}
\rput(2.5, 0){$ m_2^2$}
\psset{dotsize=3pt}
\psdots(-2.5, -2)(2.5, -2)(0, 2)
\end{pspicture}
\end{center}
\caption{\label{j3diagram}One-loop 
triangle diagrams.}
\end{figure}
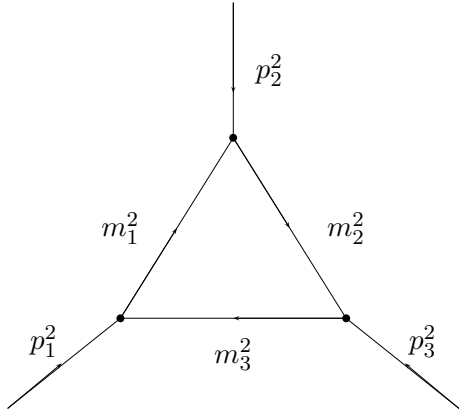

We arrive at notations for 
the calculations in this subsection.
Feynman integrals of scalar 
one-loop three-point functions 
are defined:  
\begin{eqnarray}
\label{feynmanintegral}
&&\hspace{-1.5cm}
J_3(d; \{\nu_1,\nu_2,\nu_3\})
\equiv J_3(d; \{\nu_1,\nu_2,\nu_3\}
;p_1^2,p_2^2,p_3^2;
m_1^2,m_2^2,m_3^2) =\\
&=& \int \frac{d^d k}{i\pi^{d/2}} 
\dfrac{1}{[(k+q_1)^2 -m_2^2 + i\rho]^{\nu_1}
[(k+q_2)^2 -m_3^2 + i\rho]^{\nu_2}
[(k+q_3)^2 -m_1^2 + i\rho]^{\nu_3}}. \nonumber
\end{eqnarray}
We refer hereafter 
$J_3 \equiv J_3(d; \{1,1,1\})$.
In this definition, 
the term $i\rho$ is Feynman's 
prescription and $d$ is space-time dimension. 
The internal (loop) momentum is $k$ and the 
external momenta are $p_1$, $p_2$, $p_3$. 
They are inward as described in 
Fig.~\ref{j3diagram}. We use the momenta
$q_1 = p_1$, $q_2 =p_1+p_2$ and  
$q_3 = p_1+p_2+p_3 =0$ following 
momentum conservation.
The internal masses are $m_1$, $m_2$ and 
$m_3$. $J_3$ is a function of 
$p_1^2$, $p_2^2$, $p_3^2$ and $m_1^2$, $m_2^2$, 
$m_3^2$.

It has known that an algebraically compact 
expression and numerically stable 
representation for Feynman diagrams
can be obtained by using kinematic variables such as 
the determinants of Caylay and Gram 
matrices~\cite{kajantie}. The expression
also reflects the symmetry of the corresponding 
topologies. 
We hence review the 
mentioned kinematic variables 
in the following paragraphs. 
The determinant of Cayley 
matrix of one-loop
triangle diagrams is given
\begin{eqnarray}
\label{deltan}
S_3 &=&
\left|
\begin{array}{ccc}
2m_1^2& -p_1^2 +m_1^2 +m_2^2 & -p_3^2 +m_1^2 +m_3^2   \\
-p_1^2 +m_1^2 +m_2^2 & 2m_2^2 & -p_2^2 +m_2^2 +m_3^2  \\
-p_3^2 +m_1^2 +m_3^2 & -p_2^2 +m_2^2 +m_3^2 & 2 m_3^2 \\
\end{array}
\right|.
\end{eqnarray}
In the same manner, the Cayley determinants 
of one-loop two-point Feynman 
diagrams which are obtained by shrinking an 
propagator in the three-point integrals. 
The determinants are written explicitly as:
\begin{eqnarray}
S_{12}&=&
\left|
\begin{array}{cc}
2m_1^2& -p_1^2 +m_1^2 +m_2^2 \\
-p_1^2 +m_1^2 +m_2^2 & 2m_2^2 \\
\end{array}
         \right|  = 
-\lambda(p_1^2, m_1^2,m_2^2),  \\
S_{13}&=& \left|
\begin{array}{cc}
2m_1^2 & -p_3^2 +m_1^2 +m_3^2 \\
-p_3^2 +m_1^2 +m_3^2  & 2 m_3^2 \\
\end{array}
\right| 
=-\lambda(p_3^2, m_1^2,m_3^2), \\
S_{23}
&=&\left|
\begin{array}{cc}
 2m_2^2 & -p_2^2 +m_2^2 +m_3^2 \\
 -p_2^2 +m_2^2 +m_3^2 & 2 m_3^2 \\
\end{array}
\right| 
=-\lambda(p_2^2, m_2^2,m_3^2).
\end{eqnarray}
Where $ \lambda(x, y,z) 
= x^2 +y^2 +z^2 -2 xy -2 xz -2 yz$ 
is so-called K\"allen function. 
Next, the determinant of Gram matrix 
of one-loop three-point functions is 
given
\begin{eqnarray}
 G_3 = -8  
 \left|
\begin{array}{cc}
 p_1^2    & p_1p_2 \\
 p_1p_2   & p_2^2
\end{array}
  \right|  
 = 2 \lambda(p_1^2,p_3^2,p_2^2).
\end{eqnarray}
In the same above definitions, we also get  
the Gram determinants of two-point functions
as follows
\begin{eqnarray}
 G_{12} = - 4p_1^2, \quad G_{13} 
 = - 4p_3^2, \quad G_{23} = -4p_2^2.
\end{eqnarray}
In this work, the analytic formulas
for scalar one-loop three-point integrals 
are expressed as functions with 
arguments of the ratio of the above kinematic 
determinants. Therefore, it is worth to 
introduce the following index
variables
\begin{eqnarray}
M_{3} &=&
-\dfrac{S_3}{G_3}, \quad \text{for}
\quad G_3 \neq 0,                                 \\
M_{ij} &=& 
-\dfrac{S_{ij}}{G_{ij}}, \quad \text{for}
\quad G_{ij} \neq 0, \quad \text{with}
\quad i,j =1,2,3. 
\end{eqnarray}

By introducing Feynman parameters, we then integrate 
over the loop-momentum. The resulting integral after
taking over one of Feynman parameters reads 
\begin{eqnarray}
\label{feynj3}
\dfrac{J_3}{\Gamma\left (\frac{6-d}{2} \right)} 
&=& -
\int \limits_0^1 dx\int \limits_0^{1-x} dy
\dfrac{1}{(Ax^2 + By^2+ 2C xy+ Dx + Ey + F 
- i\rho)^{\frac{6-d}{2}} }. 
\end{eqnarray}
The corresponding coefficients 
$A,B,C,\;\cdots,F$ are shown:
\begin{eqnarray}
A &=& p_1^2,\quad\quad\quad \quad    
\quad \quad \quad \quad   D \;=\;-(p_1^2+m_1^2-m_2^2), \\
B &=& p_3^2, \quad   \quad \quad 
\quad \quad \quad   \quad    
\quad E \;=\; -(p_3^2+m_1^2-m_3^2),\\ 
C &=& -p_1p_3,\quad   \quad    
\quad \quad   \quad    \quad\; F \;=\; m_1^2.
\end{eqnarray}
Detailed calculation for $J_3$ in (\ref{feynj3}) 
is presented in next subsections.
\subsection{Two light-like momenta}
We first consider the simple case which is
two light-like external momenta. Without loss
of the generality, we can take
$p_1^2 =0,~p_3^2 =0$.  
The Feynman parameter integral 
in Eq.~(\ref{feynj3}) 
is now casted into the simpler form
\begin{eqnarray}
\label{twolight-like}
J_3
&=& 
-\Gamma\left (\frac{6-d}{2} \right)
\int \limits_0^1 dx 
\int \limits_0^{1-x}dy 
\dfrac{1}
{(2Cxy + Dx + Ey+ F 
- i\rho)^{\frac{6-d}{2}} }.
\end{eqnarray}
It finds that the 
denominator function of 
the integrand in (\ref{twolight-like})
depends linearly on both $x$ and $y$. Hence, 
the integral can be taken 
easily. The resulting integral
reads after performing the $y$-integration
\begin{eqnarray}
\label{twolight-like1}
\dfrac{J_3 }
{\Gamma\left (\frac{4-d}{2} \right)} 
&=& 
\int \limits_0^1 dx \; 
\dfrac{\left[ (m_2^2-m_1^2)x 
+ m_1^2 -i\rho 
\right]^{\frac{d-4}{2}} 
-
\left[\; p_2^2x^2 
-(p_2^2 +m_3^2 -m_2^2)x 
+m_3^2 -i\rho 
\right]^{\frac{d-4}{2} }
} 
{ p_2^2 x - m_3^2 +m_1^2 }.
\nonumber\\
\end{eqnarray}
Both the integrands are singularity at 
$x_1 = (m_3^2-m_1^2)/p_2^2$. 
However, it is verified that 
\begin{eqnarray}
p_2^2x_1^2 -(p_2^2 +m_3^2 -m_2^2)x_1 + m_3^2
= (m_2^2-m_1^2) x_1  + m_1^2 =M_{3}. 
\end{eqnarray}
It means that
the residue contributions at this pole
of two integrations in Eq.~(\ref{twolight-like1}) 
will be cancelled. 
As a result, $J_3$ stays finite at this point. 
The first integral in Eq.~(\ref{twolight-like1}) 
can be formulated by mean of Appell $F_1$ functions.
For the second integral in
Eq.~(\ref{twolight-like1}), 
we can apply the formula for master integral 
as Eq.~(\ref{K2}). The result for $J_3$ then 
reads 
\begin{eqnarray}
\label{twolightlike1}
\dfrac{J_3}{\Gamma\left(\frac{4-d}{2}\right)}
&=& -\dfrac{ (m_1^2)^{\frac{d-4}{2}} }{m_3^2-m_1^2} 
\; F_1 \left( 1; 1,\frac{4-d}{2}; 2; 
\frac{p_2^2}{m_3^2-m_1^2 },  
\frac{m_1^2-m_2^2 +i\rho}{m_1^2} \right)         \\
&&\hspace{-1.8cm} +
\left(\dfrac{\partial_1 S_3}{p_2^2\;G_{23}} \right)
\Bigg[
\left(\dfrac{\partial_2 S_{23}}{G_{23}} \right) 
\dfrac{(m_3^2)^{\frac{d-4}{2} } }{M_{3} -m_3^2}
\; F_1\left(1; 1,\frac{4-d}{2}; \frac{3}{2}; 
\dfrac{M_{23} -m_3^2}{ M_{3} -m_3^2}, 
1-\frac{M_{23}}{m_3^2}  \right)\nonumber\\
&&\hspace{0cm} + 
\left( \dfrac{\partial_3 S_{23}}{G_{23} } \right)
\dfrac{(m_2^2)^{\frac{d-4}{2} } }{M_{3} -m_2^2}
\; F_1\left(1; 1, \frac{4-d}{2}; \frac{3}{2}; 
\dfrac{M_{23} -m_2^2}{ M_{3} -m_2^2}, 
 1-\frac{M_{23}}{m_2^2}  \right)   \Bigg] \nonumber \\
&&\hspace{-1.8cm} + 
\Bigg[
\dfrac{M_{23} - m_3^2}{2p_2^2(M_{3} -m_3^2)} 
\left(m^2_{3} \right)^{\frac{d-4}{2} } 
\; F_1\left(1; 1, \frac{4-d}{2}; 2; 
\dfrac{M_{23} -m_3^2}{ M_{3} -m_3^2}, 
1-\frac{M_{23}}{m_3^2}  \right)    \nonumber\\
&&\hspace{-1.3cm} 
-  \dfrac{M_{23} - m_2^2}{2p_2^2(M_{3} -m_2^2)} 
\left(m^2_{2} \right)^{\frac{d-4}{2}} 
\; F_1\left(1; 1,\frac{4-d}{2}; 2; 
\dfrac{M_{23} -m_2^2}{ M_{3} -m_2^2}, 
1-\frac{M_{23}}{m_2^2}  \right)   \Bigg]. 
\nonumber
\end{eqnarray}
Another representation for $J_3$ can be 
obtained by using Eq.~(\ref{K1})
\begin{eqnarray}
\label{twolightlike2}
 \dfrac{J_3}{\Gamma\left(\frac{4-d}{2}\right)} &=&
 -\dfrac{ (m_1^2)^{\frac{d-4}{2}} }{m_3^2-m_1^2} 
 \; F_1 \left( 1; 1, \frac{4-d}{2}; 2; 
 \frac{p_2^2}{m_3^2-m_1^2 },  
 \frac{m_1^2-m_2^2 +i\rho}{m_1^2} \right) 
 \\ 
&&\hspace{-1.8cm} +
\left( \dfrac{\partial_1 S_3}{p_2^2\;G_{23}} \right)
\Bigg[
\left( \dfrac{\partial_2 S_{23}}{G_{23}} \right) 
\dfrac{(M_{23} -i\rho)^{\frac{d-4}{2}} }{M_{3} -M_{23}}
\; F_1\left(\frac{1}{2}; 1, \frac{4-d}{2}; \frac{3}{2}; 
\dfrac{M_{23} -m_3^2}{ M_{23} -M_{3} }, 
1-\frac{m_3^2}{M_{23}}  \right)\nonumber\\
&&\hspace{0.1cm} 
+ \left( \dfrac{\partial_3 S_{23}}{G_{23} } \right)
\dfrac{(M_{23}-i\rho)^{\frac{d-4}{2}} }{M_{3} -M_{23}}
\; F_1\left(1; 1, \frac{4-d}{2}; \frac{3}{2}; 
\dfrac{M_{23} -m_2^2}{ M_{23} -M_{3}}, 
1-\frac{m_2^2}{M_{23}}  \right)  \Bigg] \nonumber \\
&&\hspace{-1.8cm} 
+ \Bigg[
\dfrac{ \left(M_{23} - i\rho\right)^{\frac{d-4}{2}} }
{2(M_{3} -M_{23} )}  
\; F_1\left(1; 1, \frac{4-d}{2}; 2; \dfrac{M_{23} -m_3^2}
{ M_{3} -M_{23} }, 
1-\frac{m_3^2}{M_{23}}  \right)    
\nonumber\\
&&\hspace{0cm}- 
\dfrac{\left(M_{23}-i\rho \right)^{\frac{d}{2} -2} }
{2p_2^2(M_{3} -M_{23} )}  
\; F_1\left(1; 1, \frac{4-d}{2}; 2; \dfrac{M_{23} -m_2^2}
{M_{3} -M_{23} }, 
1-\frac{m_2^2}{M_{23}}  \right)   \Bigg]. \nonumber
\end{eqnarray}
The results are shown in 
Eqs.~(\ref{twolightlike1}, \ref{twolightlike2}) 
are new hypergeometric representations for 
scalar one-loop three-point functions in general 
space-time dimension. In the next paragraphs, 
we consider $J_3$ in several special cases.
\begin{enumerate}
\item $\underline{m_1^2 = m_2^2 = m_3^2=0:}$ \\
One-loop triangle 
diagrams with all massless 
internal lines are considered. 
In this case, the integral $J_3$ 
in (\ref{twolight-like}) becomes
\begin{eqnarray}
J_3 &=& -\Gamma\left(\frac{6-d}{2} \right) 
\int \limits_0^1 dx  
\int \limits_0^{1-x} dy
\dfrac{1}{\left(-p_2^2\; x\; y 
-i\rho \right)^{\frac{6-d}{2} }} 
\\
&=&
\dfrac{\Gamma\left(\frac{4-d}{2}  \right) 
\Gamma\left(\frac{d-4}{2}\right)
\Gamma\left(\frac{d-2}{2}\right) }
{\Gamma(d-3)} \left(-p_2^2 -i\rho  
\right)^{\frac{d-6}{2}}. \label{case11}
\end{eqnarray}
It is important to note that we use 
$-p_2^2 x y -i\rho \equiv (-p_2^2-i\rho ) x y $. 
Therefore, for analytic continuation the result
when $p_2^2>0$, we should keep $i\rho$-term 
together with external momentum 
like $p_2^2 \rightarrow p_2^2 +i\rho$. 
The result in (\ref{case11}) shows 
full agreement with Eq.~$(4.5)$ in 
\cite{Ellis:2007qk}. 
\\
\item $\underline{m_2^2 = m_3^2 =0, m_1^2=m^2:}$ \\
If internal mass configuration 
takes $m_2^2 = m_3^2 =0, m_1^2=m^2$, $J_3$ becomes
\begin{eqnarray}
\dfrac{J_3}{\Gamma\left(\frac{4-d}{2} \right)}
&=& 
\int \limits_0^1 dx \; 
\dfrac{[m^2(1-x)]^{ \frac{d-4}{2}}}{p_2^2 x + m^2}
-\int \limits_0^1 dx \; 
\dfrac{[p^2_2x^2 - p_2^2 x -i\rho]
^{ \frac{d-4}{2} }}{p_2^2 x + m^2} \\
&&\hspace{-2.2cm}
=
\dfrac{\Gamma\left(\frac{d-2}{2}\right)}
{\Gamma\left( \frac{d}{2} \right)}  
(m^2)^{ \frac{d-6}{2}} 
\Fh21\Fz{1, 1}{\frac{d}{2} }{\dfrac{-p_2^2}{m^2}}
-\dfrac{\Gamma\left(\frac{d-2}{2}\right)^2 }
{\Gamma(d-2)} 
\dfrac{(-p_2^2 -i\rho)^{\frac{d-4}{2}} }{m^2}
\Fh21\Fz{1, \frac{d-2}{2}}{d-2}{\dfrac{-p_2^2}{m^2}}. 
\nonumber \\
\end{eqnarray}
This result is in agreement with 
Eq.~($B2$) in Ref.~\cite{Abreu:2015zaa}. 
In the limit of $p_2^2 \rightarrow  -m^2<0$, 
we arrive at
\begin{eqnarray}
\dfrac{J_3}{\Gamma\left(\frac{4-d}{2} \right)}
&=& -\left\{ 
\dfrac{\Gamma\left( \frac{d-2}{2}\right)
\Gamma\left(\frac{d-4}{2} \right) }
{\Gamma\left(d-3\right) }
-\dfrac{\Gamma\left(\frac{d-4}{2}\right) }
 {\Gamma\left( \frac{d-2}{2}\right) }
\right\}(m^2)^{ \frac{d-6}{2}}.
\end{eqnarray}

When $\mathcal{R}$e$\left(d-6\right)>0$ 
and $m^2 \rightarrow 0$, 
$J_3 \rightarrow 0$.\\
\item $\underline{m_1^2 = m_3^2:}$\\%
We are also interested in the case of 
$m_1^2 = m_3^2 $.
In such case, one should exchange the 
order of integration  
in  Eq.~(\ref{twolight-like}).  
The $x$-integration can be taken first.
Subsequently, we arrive at  
\begin{eqnarray}
\dfrac{J_3}{\Gamma\left(\frac{4-d}{2} \right)} 
&=&
-\int \limits_0^1 dy \;
\dfrac{[p^2_2y^2 - (p_2^2+m_2^2-m_1^2 )y 
+m_2^2 -i\rho]^{ \frac{d-4}{2}}
-(m_1^2)^{ \frac{d-4}{2}}
}
 {p_2^2 y -m_2^2 + m_1^2}. 
\end{eqnarray}
It is easy to find out that 
\begin{eqnarray}
\left(p^2_2y^2 
- (p_2^2+m_2^2-m_1^2 )y 
+m_2^2 \right)|_{y=(m_2^2 -m_1^2)/p_2^2}
= m_1^2.
\end{eqnarray}
As previous explanation, 
the integral $J_3$ in this case also stays 
finite at $y=(m_2^2 -m_1^2)/p_2^2$. 
Using the analytical solution for master 
integral as Eq.~(\ref{K2}) 
in appendix $A$, the result reads
\begin{eqnarray}
\dfrac{J_3}{\Gamma\left(\frac{4-d}{2} \right)  } 
&=& -\dfrac{(m_1^2)^{ \frac{d-4}{2} }}{m_2^2-m_1^2}
  \Fh21\Fz{1, 1}{2}{-\dfrac{p_2^2}{m_2^2-m_1^2}}     \\
&&\hspace{-1.8cm} +
\left( \dfrac{\partial_3 S_3}{p_2^2\; G_{23}} \right)
\Bigg[
\left( \dfrac{\partial_3 S_{23}}{G_{23}} \right) 
\dfrac{(m_2^2)^{\frac{d-4}{2}} }{M_{3} -m_2^2}
\; F_1\left(1; 1, \frac{4-d}{2}; \frac{3}{2}; 
\dfrac{M_{23} -m_2^2}{ M_{3} -m_2^2}, 
1-\frac{M_{23}}{m_2^2}  \right)\nonumber\\
&&\hspace{0cm} + 
\left(\dfrac{\partial_2 S_{23}}{G_{23} } \right)
\dfrac{(m_1^2)^{\frac{d-4}{2}} }{M_{3} -m_1^2}
\; F_1\left(1; 1,\frac{4-d}{2}; \frac{3}{2}; 
\dfrac{M_{23} -m_1^2}{ M_{3} -m_1^2}, 
1-\frac{M_{23}}{m_2^2}  \right) \Bigg] 
\nonumber \\
&&\hspace{-1.8cm} +
\Bigg[
\dfrac{M_{23} - m_2^2}{2p_2^2(M_{3} -m_2^2)} 
\left(m^2_{2} \right)^{\frac{d-4}{2}} 
\; F_1\left(1; 1, \frac{4-d}{2}; 2; 
\dfrac{M_{23} -m_2^2}{ M_{3} -m_2^2}, 
1-\frac{M_{23}}{m_2^2}  \right) \nonumber\\
&&\hspace{-0.5cm}
-\dfrac{M_{23} - m_1^2}{2p_2^2(M_{3} -m_1^2)} 
\left(m^2_{1} \right)^{\frac{d-4}{2} } 
\; F_1\left(1; 1,\frac{4-d}{2}; 2; 
\dfrac{M_{23} -m_1^2}{ M_{3} -m_1^2}, 
1-\frac{M_{23}}{m_1^2}  \right)   \Bigg],
\nonumber
\end{eqnarray}
provided that 
$\left|\frac{M_{23} -m_{1,2}^2}
{ M_{3} -m_{1,2}^2} \right|<1$,
$\left| 1-\frac{M_{23}}{m_{1,2}^2} \right|<1$.\\
\item $\underline{m_1^2 = m_2^2=0, m_3^2 =m^2, 
p_2^2 =q^2:}$\\  
This case has been calculated 
in~\cite{Abreu:2015zaa}. In this  
particular configuration, $J_3$ becomes
\begin{eqnarray}
\dfrac{J_3}{\Gamma\left (\frac{4-d}{2} \right)} 
&=& 
- \int \limits_0^1 dx
\; \dfrac{\left[ q^2x^2 -(q^2 +m^2)x +m^2 
-i\rho  \right]^{\frac{d-4}{2}}}
{q^2 x - m^2 }.
\end{eqnarray}
It is confirmed that
\begin{eqnarray}
q^2x^2 
- (q^2 +m^2)x 
+ m^2|_{x= m^2/q^2}=0.
\end{eqnarray}
As a matter of this fact, $J_3$ stays
finite at $x= m^2/q^2$. 
By shifting $x \rightarrow 1-x$, 
the resulting integral reads 
\begin{eqnarray}
\dfrac{J_3}
{\Gamma\left (\frac{4-d}{2} \right)} 
&=& 
\left(m^2-q^2-i\rho\right)^{\frac{d-6}{2} }
\int \limits_0^1 dx \; x^{\frac{d-4}{2} }
\left(1-\dfrac{q^2}{q^2 -m^2}x\right)^{\frac{d-4}{2}}.
\end{eqnarray}   
Following Eq.~(\ref{gauss-int}) in appendix $B$,
we derive this integral in terms of Gauss 
hypergeometric functions
\begin{eqnarray}
\dfrac{J_3}{\Gamma\left (\frac{4-d}{2} \right)}  
&=&\dfrac{\Gamma\left(\frac{d-2}{2}\right)  }
{\Gamma\left(\frac{d}{2}\right)}
\left(m^2-q^2-i\rho \right)^{\frac{d-6}{2}}\; 
\Fh21\Fz{\frac{6-d}{2},\frac{d-2}{2} }
{\frac{d}{2}} {\dfrac{q^2}{q^2 - m^2}} 
\nonumber\\
&=&\dfrac{\Gamma\left(\frac{d-2}{2}\right) }
{\Gamma\left(\frac{d}{2}\right)}
\left(m^2\right)^{\frac{d-6}{2}}\; 
\Fh21\Fz{1,\frac{6-d}{2}}{ \frac{d}{2}}
{\dfrac{q^2}{ m^2}},
\end{eqnarray}
provided that $\left|\frac{q^2}{m^2} \right|<1$ 
and $\mathcal{R}$e$\left(d-2\right) >0$. 
This gives full agreement result with Eq.~($B7$) in  
Ref~\cite{Abreu:2015zaa}. In the limit of 
$q^2 \rightarrow m^2$, one arrives at
\begin{eqnarray}
\dfrac{J_3}{\Gamma\left (\frac{4-d}{2} \right) }  
& = &\dfrac{\Gamma\left(d-4\right)  }
{ \Gamma\left(d-3\right)}\; 
\left( m^2 \right)^{\frac{d-6}{2}},
\end{eqnarray}
provided that $\mathcal{R}$e$\left(d-4\right) >0$. 
In additional, 
if $\mathcal{R}$e$\left(d-6\right)>0$ 
and $m^2 \rightarrow 0$, 
$J_3 \rightarrow 0$.\\
\item $\underline{m_1^2 = m_2^2 = m_3^2 =m^2:}$
\\
This case has been performed in 
Ref~\cite{Davydychev:2003mv}. 
One confirms that $M_{3} = m^2$ 
in this kinematic configuration. 
We derive analytic formula 
for $J_3$ as follows
\begin{eqnarray}
\label{rijkM}
\dfrac{J_3}{\Gamma\left (\frac{6-d}{2} \right)} 
= -\int \limits_0^1 dx\; \int \limits_0^{1-x} dy\; 
\dfrac{1}{( p_2^2 \;x\;y+m^2 - i\rho)^{\frac{6-d}{2}} }.
\end{eqnarray}
Using Mellin-Barnes relation one has 
\begin{eqnarray}
J_3  = -\frac{1}{2\pi i} 
\int\limits_{-i \infty}^{i\infty} ds \;
\Gamma\left(-s\right) \Gamma\left(\frac{6-d+2s}{2}\right) 
\int\limits_0^1 dx\; \int \limits_0^{1-x} dy\; 
\dfrac{[ (p_2^2 -i\rho) \;x\;y]^s }{ (m^2)^{3-\frac{d}{2} +s} }.
\end{eqnarray}
After taking over $x,y$-integrations,
the contour integral is taken the form of
\begin{eqnarray}
J_3 &=& - 
\frac{\sqrt{\pi}}{2\pi i} 
~(m^2)^{\frac{d-6}{2}}
\int\limits_{-i \infty}^{i\infty} ds \;
\dfrac{ \Gamma\left(-s\right) 
\Gamma\left(\frac{6-d+2s}{2}\right) 
\Gamma(s+1)^2}{4\;\Gamma(s+2)
\Gamma\left(s+\frac{3}{2} \right)} 
\left(- \dfrac{p_2^2 -i\rho}{4m^2}\right)^s. 
\nonumber\\
\end{eqnarray}
By closing the integration contour to
the right side of imaginary axis
in the $s$-complex plane, we then take 
into account the residua of sequence poles 
from $\Gamma(-s)$. The result
is presented in terms of series of 
generalized hypergeometric function
\begin{eqnarray}
\dfrac{J_3}{\Gamma\left (\frac{6-d}{2} \right)}
&=&-\dfrac{(m^2)^{\frac{d-6}{2} }}{2}\; 
\Fh32\Fz{1,1,\frac{6-d}{2} }{ 2, 
\frac{3}{2}}{ \dfrac{p_2^2 -i\rho}{4m^2}},
\end{eqnarray}
provided that 
$\left|\frac{p_2^2 -i\rho}{4m^2} \right|<1$.  
This result is in 
agreement with Ref~\cite{Davydychev:2003mv}. 
\end{enumerate}
\subsection{One light-like momentum}
We are going to proceed the method for one 
light-like momentum case. Without any loss of 
the generality, we can choose $p_3^2=0$. 
Let us use $q^2 = -2p_1p_3$. 
Applying the same previous procedure, 
one obtains Feynman parameter 
integral 
\begin{eqnarray}
\label{onelightlike}
\dfrac{J_3}{\Gamma\left (\frac{4-d}{2} \right) } 
&=&
\int \limits_0^1 dx\; 
\dfrac{\left[p_2^2x^2 -(p_2^2 +m_3^2 -m_2^2)x 
+m_3^2 -i\rho \right]^{\frac{d-4}{2} }} 
{ q^2 x + m_3^2 -m_1^2 } \\
&&- 
\int \limits_0^1 dx\; 
\dfrac{\left[p_1^2x^2 -(p_1^2 +m_1^2 -m_2^2)x 
+ m_1^2-i\rho  \right]^{\frac{d-4}{2}} }
{ q^2 x + m_3^2 -m_1^2 }.  \nonumber
\end{eqnarray}
We find that two integrands have same 
singularity pole at $x= (m_1^2-m_3^2)/q^2$. 
It is easy to verify that 
the residue contributions from this pole will 
be cancelled out. As a result, $J_3$ stays finite 
at this point. The analytic result for $J_3$ 
can be presented as a compact form:
\begin{eqnarray}
\dfrac{J_3}{\Gamma\left(\frac{4-d}{2} \right)} 
=  J_{123} +J_{231}.
\end{eqnarray}
Where the terms $J_{123}$, $J_{231}$ are 
obtained by using Eq.~(\ref{K2}). 
These terms are written in terms of
$F_1$ as follows
\begin{eqnarray}
\label{onelightlike1}
J_{123} &=& 
-\dfrac{\partial_3 S_3}{4(p_2^2-p_1^2)^2} 
\left(\dfrac{\partial_2S_{12}}{G_{12}} \right)
\dfrac{(m_1^2)^{\frac{d-4}{2} }}{M_{3} -m_1^2} 
\; F_1\left(1; 1, \frac{4-d}{2}; \frac{3}{2}; 
\dfrac{M_{12} -m_1^2}{ M_{3} -m_1^2}, 
1-\frac{M_{12}}{m_1^2}  \right)  \nonumber\\
&&-\dfrac{\partial_3 S_3}{4(p_2^2-p_1^2)^2} 
\left(\dfrac{\partial_1S_{12}}{G_{12}} \right)
\dfrac{(m_2^2)^{\frac{d-4}{2} } }{M_{3} -m_2^2} 
\; F_1\left(1; 1,\frac{4-d}{2}; \frac{3}{2}; 
\dfrac{M_{12} -m_2^2}{ M_{3} -m_2^2}, 
1-\frac{M_{12}}{m_2^2}  \right)  
\nonumber\\       
&& +\left(\dfrac{M_{12} -m_1^2 }{2(M_{3} -m_1^2)} \right) 
\dfrac{(m_1^2)^{\frac{d-4}{2}}  }{p_1^2-p_2^2}
\; F_1\left(1; 1, \frac{4-d}{2}; 2; 
\dfrac{M_{12} -m_1^2}{ M_{3} -m_1^2}, 
1-\frac{M_{12}}{m_1^2}  \right)  \nonumber\\
&&-\left(\dfrac{M_{12} -m_2^2 }{2(M_{3} -m_2^2) }\right) 
\dfrac{(m_2^2)^{\frac{d-4}{2} }}{p_1^2-p_2^2}  
\; F_1\left(1; 1, \frac{4-d}{2}; 2; 
\dfrac{M_{12} -m_2^2}{ M_{3} -m_2^2}, 
1-\frac{M_{12}}{m_2^2}  \right),
\end{eqnarray}
and 
\begin{eqnarray}
\label{onelightlike12}
J_{231} &=&
-\dfrac{\partial_1 S_3}{4(p_2^2-p_1^2)^2} 
\left(\dfrac{\partial_2S_{23}}{G_{23}} \right)
\dfrac{(m_3^2)^{\frac{d-4}{2} } }{M_{3} -m_3^2} 
\; F_1\left(1; 1, \frac{4-d}{2}; \frac{3}{2}; 
\dfrac{M_{23} -m_3^2}{ M_{3} -m_3^2}, 
1-\frac{M_{23}}{m_3^2}  \right)  \nonumber\\
&&-\dfrac{\partial_1 S_3}{4(p_2^2-p_1^2)^2} 
\left(\dfrac{\partial_3S_{23}}{G_{23}} \right)
\dfrac{(m_2^2)^{\frac{d-4}{2} } }{M_{3} -m_2^2} 
\; F_1\left(1; 1, \frac{4-d}{2}; \frac{3}{2}; 
\dfrac{M_{23} -m_2^2}{ M_{3} -m_2^2}, 
1-\frac{M_{23}}{m_2^2}  \right)  
\nonumber\\       
&& +\left(\dfrac{M_{23} -m_3^2 }{2(M_{3} -m_3^2)} \right) 
\dfrac{(m_3^2)^{\frac{d-4}{2} }}{p_2^2-p_1^2}  
\; F_1\left(1; 1, \frac{4-d}{2}; 2; 
\dfrac{M_{23} -m_3^2}{ M_{3} -m_3^2}, 
1-\frac{M_{23}}{m_3^2}  \right)  \nonumber\\
&&-\left(\dfrac{M_{23} -m_2^2 }{2(M_{3} -m_2^2)}\right) 
\dfrac{(m_2^2)^{\frac{d-4}{2} }}{ p_2^2-p_1^2}  
\; F_1\left(1; 1,\frac{4-d}{2}; 2; 
\dfrac{M_{23} -m_2^2}{ M_{3} -m_2^2}, 
1-\frac{M_{23}}{m_2^2}  \right).
\end{eqnarray}
It is important to note that the result for $J_3$ 
in (\ref{onelightlike1}, \ref{onelightlike12})
are only valid if  the 
absolute value of  the arguments of 
{the Appell $F_1$ functions} in this presentation 
are less  than $1$. If these kinematic 
variables do not meet this condition. 
One has to perform analytic {continuations} 
for Appell $F_1$ 
functions~\cite{olsson}.

We can get another representation for
$J_{123}, J_{231}$ by using a transformation 
for $F_1$ (seen appendix $B$). Taking $J_{123}$
as a example, one has
\begin{eqnarray}
\label{onelightlike2}
J_{123}&=& 
\dfrac{\partial_3 S_3}{4(p_2^2-p_1^2)^2} 
\left(\dfrac{\partial_2S_{12}}{G_{12}} \right)
\dfrac{(M_{12}-i\rho)^{\frac{d-4}{2} } }{M_{12}-M_{3}} 
\; F_1\left(\frac{1}{2}; 1,\frac{4-d}{2}; 
\frac{3}{2}; \dfrac{M_{12} -m_1^2}{ M_{12} -M_{3}}, 
1-\frac{m_1^2}{M_{12}}  \right)  \nonumber\\
&&\hspace{-0.5cm}
+\dfrac{\partial_3 S_3}{4(p_2^2-p_1^2)^2} 
\left(\dfrac{\partial_1 S_{12}}{G_{12}} \right)
\dfrac{(M_{12} -i\rho )^{\frac{d-4}{2} } }{M_{12} -M_{3}} 
\; F_1\left(\frac{1}{2}; 1,\frac{4-d}{2}; \frac{3}{2}; 
\dfrac{M_{12} -m_2^2}{ M_{12} -M_{3} }, 
1-\frac{m_2^2}{M_{12}}  \right)  
\nonumber\\       
&& \hspace{-0.5cm}
+ \left(\dfrac{M_{12} -m_1^2 }{2(M_{3} -M_{12} )}\right) 
\dfrac{(M_{12} -i\rho)^{\frac{d-4}{2} }}{p_2^2-p_1^2} 
\; F_1\left(1; 1, \frac{4-d}{2}; 2; 
\dfrac{M_{12} -m_1^2}{ M_{12} -M_{3} }, 
1-\frac{m_1^2}{M_{12}}  \right)  \nonumber\\
&&\hspace{-0.5cm}
-\left(\dfrac{M_{12} -m_2^2 }{2(M_{3} -M_{12}) }\right) 
\dfrac{(M_{12} -i\rho)^{\frac{d-4}{2} }}{p_2^2-p_1^2}  
\; F_1\left(1; 1, \frac{4-d}{2}; 2; 
\dfrac{M_{12} -m_2^2}{ M_{12} -M_{12}}, 
1-\frac{m_2^2}{M_{12}}  \right).
\end{eqnarray}
From Eqs.~(\ref{onelightlike1}, \ref{onelightlike2}), 
we can perform analytic 
continuation the result in the limits of 
$M_{ij}= 0, m_i^2, m_j^2$ (for $i,j=1,2,3$) 
and $M_{3} =0$, etc. This can be worked out 
by applying the transformations for Appell $F_1$ 
functions, seen appendix $B$.

The results in Eqs.~(\ref{onelightlike1}, 
\ref{onelightlike2}) are new hypergeometric  
representations for scalar one-loop three-point 
functions for this case in general $d$. 
Several special cases for $J_3$ are considered
in the next paragraphs. 
\begin{enumerate}  
\item $\underline{m_1^2 = m_2^2 =m_3^2=0:}$\\
One first arrives at the case of all 
massless internal lines. In this case, 
Eq.~(\ref{onelightlike}) becomes
\begin{eqnarray}
\dfrac{J_3}{\Gamma\left(\frac{4-d}{2} \right)} 
&=& -\int \limits_0^1 dx\; 
\dfrac{\left[p_2^2x^2 
- p_2^2x -i\rho \right]^{\frac{d-4}{2} } } 
{ (p_2^2-p_1^2)x  } 
+
\int \limits_0^1 dx\;
\dfrac{\left[p_1^2x^2 
-p_1^2x -i\rho  \right]^{\frac{d-4}{2} } }
{ (p_2^2-p_1^2) x}  
\\
&=&\sum\limits_{k=1}^2
\dfrac{(-1)^k}
{p_1^2-p_2^2}  \int \limits_0^1 dx\; 
(-p_k^2 +i\rho)^{\frac{d-4}{2} } 
x^{\frac{d-6}{2}} (1-x)^{\frac{d-4}{2}} 
\\
&=&\dfrac{\Gamma\left (\frac{d-4}{2}\right) 
\Gamma\left (\frac{d-2}{2}\right)}
{ \Gamma(d-3)}\sum\limits_{k=1}^2
(-1)^k\;\dfrac{ (-p_k^2 
+i\rho)^{\frac{d-4}{2}}}{ p_1^2-p_2^2}.
\end{eqnarray} 
We have already used 
$p_2^2x^2 -p_2^2x -i\rho = (-p_2^2 -i\rho) x(1-x) $.  
This result coincides with Eq.~($4.6$) in 
Ref.~\cite{Ellis:2007qk}.  
In the limit of $p_2^2 \rightarrow p_1^2$, 
one arrives at
\begin{eqnarray}
\dfrac{J_3}{\Gamma\left(\frac{4-d}{2} \right)} 
&=&  \dfrac{
\Gamma^2\left (\frac{d-4}{2}\right)}{\Gamma(d-3)}
\; (-p_2^2 -i\rho)^{ \frac{d-6}{2}}. 
\end{eqnarray}
If $\mathcal{R}$e$\left(d-6\right)>0$ 
and $p_2^2 \rightarrow 0$,
the integral $J_3 \rightarrow 0 $.
\\
\item $\underline{m_1^2 =m_2^2 =0, m_3^2 = m^2:}$\\
We are concerning the case in which the internal masses 
have $m_1^2 =m_2^2 =0, m_3^2 = m^2$. Let us note that 
$q^2 = 2p_1p_3$, Eq.~(\ref{onelightlike}) now gets 
the form of
\begin{eqnarray}
\dfrac{J_3}{\Gamma\left(\frac{4-d}{2} \right)} 
&=& -
\int \limits_0^1 dx\; 
\dfrac{\left[p_2^2x^2 -(p_2^2 +m^2 )x +
m^2 -i\rho  \right]^{\frac{d-4}{2} }
-\left[p_1^2x^2 -p_1^2 x 
 -i\rho  \right]^{\frac{d-4}{2}}
} {q^2x - m^2 }.  
\nonumber\\
\end{eqnarray}
We make a change variable like $x \rightarrow 1-x$ 
for the first integral. Subsequently, it is 
presented in terms of Appell $F_1$ functions. 
While the second integral is formulated 
by mean of Gauss hypergeometric functions. 
The result is shown in concrete as follows
\begin{eqnarray}
\dfrac{J_3}{\Gamma\left (\frac{4-d}{2} \right)} 
&=&
-\dfrac{\Gamma\left (\frac{d-2}{2} \right)^2  }
{\Gamma\left (d-2\right)}  
\dfrac{ (-p_1^2 -i\rho)^{ \frac{d-4}{2}}}{m^2}
\Fh21\Fz{1, \frac{d-2}{2}}{d-2}{ \dfrac{q^2}{m^2} }
\\
&& 
\hspace{-1.8cm} + \dfrac{
\Gamma\left (\frac{d-2}{2} \right)}
{\Gamma\left (\frac{d}{2} \right)} 
\dfrac{(m^2-p_2^2 -i\rho)^{\frac{d-4}{2} } }
{q^2-m^2}
F_1\left(\frac{d-2}{2}; 1,\frac{4-d}{2}; \frac{d}{2}; 
\dfrac{q^2}{q^2-m^2}; \dfrac{p_2^2}{p_2^2 -m^2+i\rho} 
\right) \nonumber
\end{eqnarray}
provided that $\left|q^2/m^2 \right|<1$ and 
$\mathcal{R}$e$\left(d-2\right)>0$. 
One finds another representation for $J_3$ 
by applying Eq.~(\ref{f1relation1}) in appendix $B$
\begin{eqnarray}
\label{j300m}
\dfrac{J_3}{\Gamma\left (\frac{4-d}{2} \right)}
&=&
-\dfrac{\Gamma\left(\frac{d-2}{2} \right)^2  }
{\Gamma\left (d-2\right)} 
\dfrac{ (-p_1^2 -i\rho)^{ \frac{d-4}{2}}}{m^2}
\Fh21\Fz{1, \frac{d-2}{2}}{d-2}{ \dfrac{q^2}{m^2} }
\\
&&
 - \dfrac{
\Gamma\left (\frac{d-2}{2} \right)}
{\Gamma\left (\frac{d}{2} \right)} 
(m^2)^{\frac{d-6}{2} } \; 
F_1\left(1; 1,\frac{4-d}{2}; \frac{d}{2}; 
\dfrac{q^2}{m^2}; \dfrac{p_2^2}{m^2-i\rho} 
\right). \nonumber
\end{eqnarray}
This representation gives agreement 
result with Eq.~($C6$) in \cite{Abreu:2015zaa}. 
In the limit of $q^2 \rightarrow m^2$, one gets
\begin{eqnarray}
\dfrac{J_3}{\Gamma\left (\frac{4-d}{2} \right)}
&=&-\dfrac{\Gamma\left(\frac{d-4}{2} \right)}
{\Gamma\left (\frac{d-2}{2} \right)} 
(m^2)^{\frac{d-6}{2} } \; 
\Fh21\Fz{1,\frac{4-d}{2} }{ \frac{d-2}{2}}
{ \dfrac{p_2^2}{m^2-i\rho} }
\nonumber\\
&& 
-\dfrac{\Gamma\left (\frac{4-d}{2} \right)
\Gamma\left (\frac{d-2}{2} \right)^2
 \Gamma\left (\frac{d}{2} \right)^2 }
 {\Gamma\left (d-3\right)}  
 \dfrac{ (-p_1^2 -i\rho)^{\frac{d-4}{2}}}{m^2}.
 \end{eqnarray}
When $p_2^2 \rightarrow m^2$, the result in 
(\ref{j300m}) reads
\begin{eqnarray}
\dfrac{J_3}{\Gamma\left (\frac{4-d}{2} \right)}
&=&-\dfrac{\Gamma\left (d-3\right)}
{\Gamma\left (d-2 \right)} 
(m^2)^{\frac{d-6}{2} } 
\;\Fh21\Fz{1, 1}{d-2}{ \dfrac{q^2}{m^2} } \\
&&-\dfrac{
\Gamma\left (\frac{d-2}{2} \right)^2  }
{\Gamma\left (d-2\right)}  
\dfrac{ (-p_1^2 -i\rho)^{ \frac{d-4}{2}}   }{m^2}
\Fh21\Fz{1, \frac{d-2}{2}}{d-2}
{ \dfrac{q^2}{m^2} }. \nonumber
 \end{eqnarray}
 \item $\underline{m_1^2 = m_3^2 =0, m_2^2 =m^2:}$\\
This case has been performed 
in Ref~\cite{Fleischer:2003rm}. 
The Feynman parameter
integral for $J_3$ in this case reads
\begin{eqnarray}
\dfrac{J_3}{\Gamma\left(\frac{4-d}{2} \right)  } 
&=&\int \limits_0^1 dx\;
\dfrac{\left[p_2^2x^2 -(p_2^2 -m^2)x -i\rho  
\right]^{\frac{d-4}{2}} 
- \left[p_1^2x^2 -(p_1^2  -m^2)x -i\rho  \right]
^{\frac{d-4}{2}}  }{ -2p_1p_3 \; x }\nonumber\\
&& \\
&=&\sum\limits_{k=1}^2
(-1)^k \dfrac{(m^2 - p_k^2 -i\rho)
^{\frac{d-4}{2} }}{p_1^2 -p_2^2}\; 
\int \limits_0^1 dx\; x^{\frac{d-6}{2}}
\left[1- \dfrac{p_k^2}{p_k^2 -m^2 +i\rho}
\;x \right]^{ \frac{d-4}{2}} 
\nonumber\\
&& \\
&=& \dfrac{ \Gamma\left(\frac{d-4}{2}\right) }
{ \Gamma\left( \frac{d-2}{2}\right)} \; 
\sum\limits_{k=1}^2 (-1)^k  
\frac{(m^2 - p_k^2 -i\rho)^{\frac{d-4}{2}}}{p_1^2 -p_2^2}\; 
\Fh21\Fz{\frac{4-d}{2}; \frac{d-4}{2} }{\frac{d-2}{2}}
{\dfrac{p_k^2}{p_k^2 -m^2 +i\rho}   }
\nonumber\\
&& \\
&=&\dfrac{\Gamma\left(\frac{d-4}{2}\right) }
{ \Gamma\left( \frac{d-2}{2}\right)} \; 
\sum\limits_{k=1}^2 (-1)^k
\frac{(m^2)^{\frac{d-4}{2}}}{p_1^2 -p_2^2}\; 
\Fh21\Fz{1;\frac{4-d}{2}}
{\frac{d-2}{2}}{\dfrac{p_k^2}{m^2-i\rho}   }. 
 \end{eqnarray}
This gives a perfect agreement with 
Ref~\cite{Fleischer:2003rm}. In the limit 
of $p_2^2\rightarrow p_1^2$, one first uses
\begin{eqnarray}
\dfrac{d}{dp_2^2}
\left\{\Fh21\Fz{1;\frac{4-d}{2} }
{\frac{d-2}{2}}{\dfrac{p_2^2}{m^2-i\rho} } \right\}
= -\dfrac{1}{m^2}\dfrac{d-4}{d-2} 
\;\Fh21\Fz{2;\frac{6-d}{2} }{\frac{d}{2} }
{ \dfrac{p_2^2}{m^2-i\rho} }.
\end{eqnarray}
The result reads
\begin{eqnarray}
J_3 &=& 2 
\dfrac{\Gamma\left(\frac{4-d}{2} \right) }
{2-d} \; (m^2)^{\frac{d-6}{2} }
\Fh21\Fz{2;\frac{6-d}{2} }{\frac{d}{2}}
{\dfrac{p_2^2}{m^2-i\rho}   }. 
\end{eqnarray} 
\item  $\underline{m_2^2 = m_3^3 =0,
m_1^2=m^2:}$ 
\\
Let us note that $q^2 = 2p_1p_3$, 
the integral $J_3$ in (\ref{onelightlike}) 
is written by
 \begin{eqnarray}
\dfrac{J_3}{\Gamma\left(\frac{4-d}{2} \right)} 
&=& \int \limits_0^1 dx\; \Bigg\{ 
\dfrac{\left[p_2^2x^2 
- p_2^2 x  -i\rho  \right]^{\frac{d-4}{2}}}
{-q^2 x -m^2 } - \dfrac{\left[p_1^2x^2 -(p_1^2 -m^2)x 
-i\rho  \right]^{\frac{d-4}{2} }}
{ q^2\; x -q^2  -m^2 } \Bigg\}.  \nonumber\\
\end{eqnarray}
Here we have already performed
a shift $x \rightarrow 1-x $ for 
the second integral. 
It is then expressed in terms of 
Appell $F_1$ functions. While 
the first integral is presented in 
terms of Gauss hypergeometric functions. 
Combining all these terms, analytic
result for $J_3$ reads
\begin{eqnarray}
\dfrac{J_3}{\Gamma\left(\frac{4-d}{2} \right)} 
&=& \dfrac{\Gamma\left (\frac{d-2}{2} \right)  }
{\Gamma\left (\frac{d}{2} \right)} 
\dfrac{(m^2-p_1^2 -i\rho)^{\frac{d-4}{2} }}
{q^2+m^2} \times\\
&& \hspace{1cm}\times 
F_1\left(\dfrac{d-2}{2}; 1,\frac{4-d}{2}; \frac{d}{2}; 
\dfrac{q^2}{q^2 + m^2}; \dfrac{p_1^2}{p_1^2 -m^2+i\rho} 
 \right) \nonumber\\
&&-
\dfrac{\Gamma\left(\frac{d-2}{2} \right)^2  }
{\Gamma\left(d-2\right)}  
\dfrac{(-p_2^2 -i\rho)^{ \frac{d-4}{2} } }{m^2}
\Fh21\Fz{1, \frac{d-2}{2} }{d-2}{ -\dfrac{q^2}{m^2} }.
\nonumber
\end{eqnarray}
Another representation for $J_3$ is derived 
by using Eq.~(\ref{f1relation1}) in appendix $B$. 
It is
\begin{eqnarray}
\dfrac{J_3}{\Gamma\left (\frac{4-d}{2} \right)}
&=&\dfrac{\Gamma\left (\frac{d-2}{2} \right)  }
{\Gamma\left(\frac{d}{2} \right)}\; (m^2)^{\frac{d-6}{2} } 
F_1\left(1; 1, \frac{4-d}{2}; \frac{d}{2}; 
-\dfrac{q^2}{m^2}; \dfrac{p_1^2}{m^2-i\rho} 
 \right) \nonumber\\
&&-
 \dfrac{\Gamma\left (\frac{d-2}{2} \right)^2  }
 {\Gamma\left (d-2\right)}  
 \dfrac{ (-p_1^2 -i\rho)^{ \frac{d-4}{2} } }{m^2}
 \Fh21\Fz{1, \frac{d-2}{2} }{d-2}{ -\dfrac{q^2}{m^2} }.
 \end{eqnarray}
From this representation, one can perform 
analytic continuation this result 
in the limits of $q^2 \rightarrow -m^2$
and $p_1^2 \rightarrow m^2$. 
\\
\item $\underline{p_1^2=p_2^2 \neq 0:}$ 
\\
We are going to consider an interesting case in which is 
$p_1^2=p_2^2 \neq 0$. In such the case, one recognizes 
that $G_3=0$. We can present $J_3$ in terms of two scalar 
one-loop two-point functions as follows 
\begin{eqnarray}
\label{j3gram}
J_3 &=& -\Gamma\left (\frac{4-d}{2} \right)
\int \limits_0^1 dx\; \Bigg\{ 
\dfrac{\left[p_2^2x^2 -(p_2^2 +m_3^2 -m_2^2)x +m_3^2 -i\rho  
\right]^{\frac{d-4}{2} }}{m_3^2 -m_1^2 }   \\
&&\hspace{3cm} 
+ \dfrac{\left[p_1^2x^2 -(p_1^2 +m_1^2 -m_2^2)x +m_1^2
-i\rho \right]^{\frac{d-4}{2} }}{m_1^2 -m_3^2 }    
\Bigg\} \nonumber\\
&=&-\Gamma\left (\frac{4-d}{2}\right)
(J_{231} +J_{123}).
\end{eqnarray}
Both terms in right 
hand side of Eq.~(\ref{j3gram}) are 
determined as Feynman parameter integrals 
of scalar one-loop two-point functions. They are 
calculated in detail as follows. 
Let us consider $J_{231}$ 
as a example. We can rewrite
$J_{231}$ in  the following form: 
\begin{eqnarray}
 J_{231} &=&\dfrac{1}{m_3^2 -m_1^2} \int \limits_0^1 dx\;  
 \left[p_2^2(x -x_{23})^2 +M_{23} -i\rho \right]^{\frac{d-4}{2} }.
\end{eqnarray}
The integral will be worked out by 
applying Mellin-Barnes relation which is
\begin{eqnarray}
 J_{231} &=&
 \dfrac{2}{m_3^2 -m_1^2}\; \frac{1}{2\pi i} \int\limits_{-i\infty}^{i\infty}
 \dfrac{\Gamma(-s) \Gamma\left(\frac{4-d+2s}{2} \right) \Gamma(s+\frac{1}{2})}
 {\Gamma\left(\frac{4-d}{2} \right)\Gamma(s+\frac{3}{2})}
\\
 && \hspace{0cm}\times
 \left( M_{23} -i\rho \right)^{\frac{d-4}{2} }  
 \left[
 x_{23} \left( \dfrac{p_2^2x_{23}^2}{M_{23} -i\rho}\right)^s 
 + x_{32} \left( \dfrac{p_2^2x_{32}^2}{M_{23} -i\rho}\right)^s 
\;\; \right]           \nonumber\\
 &=& \dfrac{\left( M_{23}-i\rho\right)^{\frac{d-4}{2}} }{ m_3^2 -m_1^2}
 \left\{ 
 x_{23}\; \Fh21\Fz{\frac{4-d}{2}; \frac{1}{2}}{\frac{3}{2}}{ 
 \dfrac{p_2^2x_{23}^2}{M_{23} -i\rho} } 
 +(2 \leftrightarrow 3)
 \right\}.
\end{eqnarray}
The $J_3$ now is casted into the form of 
\begin{eqnarray}
\dfrac{J_3}{\Gamma\left (\frac{4-d}{2} \right)}
&=&\Bigg\{
\left(\dfrac{\partial_2 S_{23} }{G_{23}}\right) 
\dfrac{(M_{23} -i\rho)^{ \frac{d-4}{2}}}{m_3^2-m_1^2}
\Fh21\Fz{\frac{4-d}{2}; \frac{1}{2} }
{\frac{3}{2}}{\dfrac{M_{23}-m_3^2}{M_{23} -i\rho} }
+(2\leftrightarrow3) \Bigg\} \nonumber
\\
&&
\hspace{-0.2cm}
+\Bigg\{
\left(\dfrac{\partial_2 S_{12} }{G_{23}}\right) 
\dfrac{(M_{12} -i\rho)^{ \frac{d-4}{2} }} {m_1^2-m_3^2}
\Fh21\Fz{\frac{4-d}{2}; \frac{1}{2} }{\frac{3}{2}}
{\dfrac{M_{12}-m_1^2}{M_{12} -i\rho} }
+(2\leftrightarrow 1) \Bigg\} 
\nonumber\\
\end{eqnarray}
\end{enumerate}
provided that 
$\left|\frac{M_{12}-m_{1,3}^2}{M_{12} -i\rho}\right|<1$. 
\subsection{General case}
We are going to generalize the method 
for the general case in which
$p_i^2 \neq 0$ for $i=1,2,3$. Following 
an idea in~\cite{'tHooft:1978xw}, 
we first apply the Euler transformation 
like $y \rightarrow y - \beta x$, 
the polynomial written in terms of $x,y$ in 
$J_3$'s integrand will becomes
\begin{eqnarray}
 &&\hspace{-0.5cm} 
 A x^2+ By^2+2Cxy+ Dx+Ey+F- i\rho =\\
 &&= (B\beta^2 - 2C\beta + A) x^2 
 + B y^2 +2(C-B\beta) xy + (D - E\beta)x + E y + F -i\rho.
 \nonumber
\end{eqnarray}
By choosing $\beta$ is one of the 
roots of following equation
\begin{eqnarray}
\label{alpha}
 B \beta^2 - 2C \beta + A =0, \quad \text{or} \quad \beta= 
 \dfrac{C \pm \sqrt{C^2 - AB}}{B},
\end{eqnarray}
the integral $J_3$ is casted into 
\begin{eqnarray}
&&\dfrac{J_3}{\Gamma\left (\frac{6-d}{2} \right)} =
\int \limits_0^1 dx\int \limits_{\beta x}^{1-(1-\beta)x} dy\; 
\left\{ \Big[2(C-B\beta)y + D - E\beta \Big] x 
+ By^2+ Ey+ F - i\rho   \right\}^{\frac{d-6}{2}} .
\nonumber\\
\end{eqnarray}
It is also important to note 
that the final result will be independent of
$\beta$ in Eq.~(\ref{alpha}). It means that
we are free to choice one of roots $\beta$ 
in Eq.~(\ref{alpha}). As a result,
the integrand is linearized of $x$, 
hence the $x$-integration can be evaluated
first. In order to work out the $x$-integration, 
we split the integration as follows
\begin{eqnarray}
\int \limits_0^1 dx\; \int \limits_{\beta x}^{1-(1-\beta)x} dy
=\left\{  \int \limits_0^1 dy \; \int \limits_{0}^{1} dx 
-  \int \limits_0^{\beta}dy\; \int \limits_{\frac{y}{\beta}}^{1} dx
-  \int \limits_{\beta}^1 dy\; \int \limits_{\frac{y-1}{\beta-1}}^{1} dx
\right\}.
\end{eqnarray}
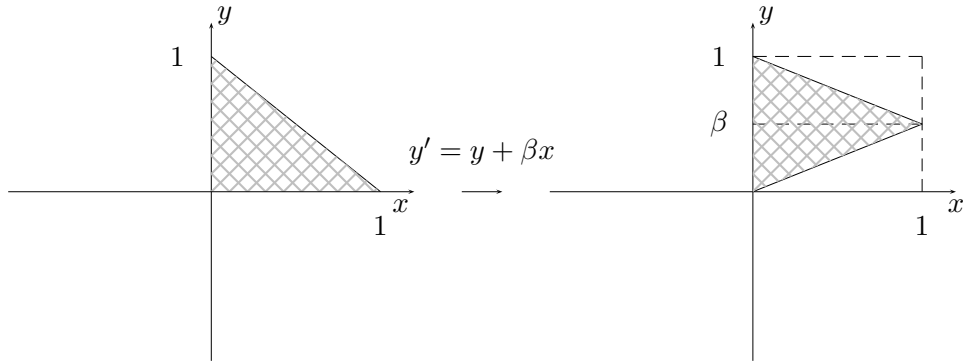
\begin{figure}[ht]
\begin{center}
\begin{pspicture}(-3, -2.5)(3, 2.5)
\psset{linewidth=.1pt}
\psset{unit = 0.9}
\psline{->}(-4,-2.5)(-4,2.5)
\psline(-4, 2)(-1.5, 0)
\rput(-4.5, 2){$1$}
\rput(-1.5, -0.5){$1$}
\pspolygon[fillstyle=crosshatch, fillcolor=gray,
linestyle=none, hatchcolor=lightgray](-4,2)(-4,0)(-1.5,0)
\psline{->}(4,-2.5)(4,2.5)
\psline{->}(-7,0)(-1,0)
\psline{->}(1,0)(7,0)
\psline(4,2)(6.5,1)
\psline(4,0)(6.5,1)
\psline[linestyle=dashed](4,1)(6.5,1)
\psline[linestyle=dashed](4,2)(6.5,2)
\psline[linestyle=dashed](6.5,0)(6.5,2)
\pspolygon[fillstyle=crosshatch, fillcolor=gray,
linestyle=none, hatchcolor=lightgray](4,2)(4,0)(6.5,1)
\rput(-3.8, 2.6){$y$}
\rput(6.5, -0.5){$1$}
\rput(3.5, 2){$1$}
\rput(3.5, 1){$\beta$}
\rput(4.2, 2.6){$y$}
\rput(-1.2,-0.2){$x$}
\rput(7,-0.2){$x$}
\rput(0., 0.6){$y' = y+ \beta x$}
\psline{->}(-0.3,0)(0.3,0)
\end{pspicture}
\end{center}
\caption{\label{region} The integration 
looks like after after shifting 
$y \rightarrow y - \beta x $.}
\end{figure}
To archive a more symmetric 
form we make a further transformations 
$y \rightarrow  \beta (1 - y )$ for 
the second integral and $y \rightarrow  1 -(1 - \beta )y$ 
for the third integral respectively. 
This brings some order in the arguments of the 
integrands. The denominators are all of 
the linear form $\sim (y - y_i)$ for $i = 1, 2, 3$. 
It is also easy to confirm that all $y_i$ follows 
the equations 
\begin{eqnarray}
 p_i^2 (y_k -x_{ij})^2 +M_{ij} -i\rho = M_3 -i\rho, 
\end{eqnarray}
for $i,j,k =1,2,3$. 

Finally, following an idea 
in~\cite{'tHooft:1978xw}, 
we add extra terms which sum 
all of them is to zero for cancelling
the residue of the pole at $y_i$.
The result reads
\begin{eqnarray}
&&\hspace{-0.5cm}
\dfrac{ \lambda^{1/2}(p_1^2, p_2^2, p_3^2) }
{\Gamma\left (\frac{4-d}{2} \right)} 
\; J_3 =  \\ 
&&=
-\int\limits_0^1 dy \frac{\left[p_1^2 y^2 
-(p_1^2 + m_1^2 -m_2^2)y +m_1^2 -i\rho \right]
^{\frac{d-4}{2} }}{y-y_3} 
+  \int\limits_0^1 dy 
\frac{(M_{3} -i\rho)^{\frac{d-4}{2} }}{y-y_3} \nonumber\\
&& \hspace{0cm} \quad
-\int\limits_0^1 dy \frac{\left[p_2^2 y^2 
-(p_2^2 + m_2^2 -m_3^2)y +m_2^2 -i\rho \right]
^{\frac{d-4}{2} }}{y-y_1}  
+  \int\limits_0^1 dy 
\frac{(M_{3} -i\rho)^{\frac{d-4}{2} }}{y-y_1}                   
\nonumber \\
&& \hspace{0cm} \quad
-\int\limits_0^1 dy \frac{\left[p_3^2 y^2 
-(p_3^2 + m_3^2 -m_1^2)y +m_3^2 -i\rho \right]
^{\frac{d-4}{2} }}{y-y_2}  
+  
\int\limits_0^1 dy 
\frac{(M_{3} -i\rho)^{\frac{d-4}{2} }}{y-y_2}. 
\nonumber
\end{eqnarray}
The analytic result for $J_3$ 
can be written in a compact form  
\begin{eqnarray}
\label{j3final}
\dfrac{J_3}{\Gamma\left (\frac{4-d}{2} \right)} 
&=& \;\left\{ 
-\dfrac{
(M_{3} -i\rho)^{\frac{d-4}{2} }
} {\lambda^{1/2}(p_1^2, p_2^2, p_3^2)}
\cdot J^{(d=4)}_{123}  + 
\dfrac{1} {\lambda^{1/2}(p_1^2, p_2^2, p_3^2)}
\cdot
J^{(d)}_{123}
\right\} \nonumber\\
 && \nonumber\\
 && +
\left\{ 
 (1,2,3) \leftrightarrow (2,3,1)
\right\}
+ 
\left\{ 
(1,2,3) \leftrightarrow (3,1,2)
\right\},
\end{eqnarray}
with
\begin{eqnarray}
J_{ijk}^{(d)}&=& - \int\limits_0^1 dy 
\frac{\left[p_i^2 y^2 -(p_i^2 + m_i^2 
-m_j^2)y +m_i^2 -i\rho \right]
^{\frac{d-4}{2} }}{y-y_k}, 
\end{eqnarray}
for $i,j,k =1,2,3.$ Where the integrand's 
poles are given
\begin{eqnarray}
y_1 &=& 1-\dfrac{D - E\beta +2(C - B\beta)}{2(1-\beta) 
(C - B\beta)}, 
\; y_2 =  1+ \dfrac{D - E\beta}{2(C - B\beta)},                                
\; y_3 = - \dfrac{D - E\beta }{2\beta(C - B\beta)}.
\end{eqnarray}
Applying the formula for master integral in 
appendix $A$, we will present 
the result of $J_{ijk}$ in terms of 
Appell $F_1$ functions.  
For instant, one takes $J_{123}$  
for an example. This term  is expressed as follows
\begin{eqnarray}
\label{j123general1}
J_{123} &=& 
-\left(\dfrac{\partial_3 S_3 }{\sqrt{8G_3 } } \right)
\Bigg[ \left(\dfrac{\partial_2 S_{12} }{G_{12} } \right)
\dfrac{ (m_1^2)^{\frac{d -4}{2} } }{M_{3}-m_1^2 } 
F_1\left(1; 1,\frac{4-d}{2}; \frac{3}{2}; 
\dfrac{M_{12} -m_1^2}{M_{3} -m_1^2}, 
1-\dfrac{M_{12} }{m_1^2} \right)\nonumber\\
&&\hspace{1.7cm}+ 
\left(\dfrac{\partial_1 S_{12} }{G_{12} } \right)
\dfrac{ (m_2^2)^{\frac{d-4}{2} }}{M_{3}-m_2^2 } 
F_1\left(1; 1, \frac{4-d}{2}; \frac{3}{2}; 
\dfrac{M_{12} -m_2^2}{M_{3} -m_2^2}, 
1-\dfrac{M_{12} }{m_2^2} \right)  \Bigg] \nonumber\\
&&+\Bigg[ \dfrac{M_{12} -m_1^2}
{2(M_{3}-m_1^2)} (m_1^2)^{\frac{d-4}{2}}
F_1\left(1; 1, \frac{4-d}{2}; 2; 
\dfrac{M_{12} -m_1^2}{M_{3} -m_1^2}, 
1-\dfrac{M_{12} }{m_1^2} \right)  
\\
&& \hspace{1.4cm}
-\dfrac{M_{12} -m_2^2}{2(M_{3}-m_2^2)} 
(m_2^2)^{\frac{d-4}{2} }
F_1\left(1; 1, \frac{4-d}{2}; 2; 
\dfrac{M_{12} -m_2^2}{M_{3} -m_2^2}, 
1-\dfrac{M_{12} }{m_2^2} \right) \;\;\Bigg], 
\nonumber
\end{eqnarray}
provided that $\left|\frac{M_{12} -m_{1;2}^2}
{M_{3} -m_{1;2}^2}\right|,
\;\left|1-\frac{M_{12} }{m_{1;2}^2}\right|<1$. 
One finds another representation for $J_{123}$ is as 
(by applying Eq.~(\ref{K1}) in appendix $A$)
\begin{eqnarray}
\label{j123general2}
J_{123} &=& 
-\left(\dfrac{\partial_3 S_3 }{\sqrt{8G_3 } } \right)
\Bigg[ \left(\dfrac{\partial_2 S_{12} }{G_{12} } \right)
\dfrac{ (M_{12} -i\rho)^{\frac{d-4}{2} } }{M_{3}-M_{12} } 
F_1\left(\frac{1}{2}; 1,\frac{4-d}{2}; 
\frac{3}{2}; \dfrac{M_{12} -m_1^2}{M_{12} -M_{3} }, 
1-\dfrac{m_1^2}{M_{12} } \right)\nonumber\\
&&\hspace{1.5cm}+ 
\left(\dfrac{\partial_1 S_{12} }{G_{12} } \right)
\dfrac{ (M_{12} -i\rho)^{\frac{d-4}{2} }}{M_{3}-M_{12} } 
F_1\left(\frac{1}{2}; 1, \frac{4-d}{2}; 
\frac{3}{2}; \dfrac{M_{12} -m_2^2}{M_{12} -M_{3} }, 
1-\dfrac{m_2^2}{M_{12} } \right)
\Bigg] 
\nonumber\\
&&+\Bigg[ \dfrac{M_{12} -m_1^2}{2(M_{12}-M_{3}) } 
(M_{12} -i\rho)^{\frac{d-4}{2} }
F_1\left(1; 1,\frac{4-d}{2}; 
2; \dfrac{M_{12} -m_1^2}{M_{12} -M_{3} }, 
1-\dfrac{m_1^2}{M_{12} } \right)                
\\
&& \hspace{1.0cm}
-\dfrac{M_{12} -m_2^2}{2(M_{12}-M_{3}) } 
(M_{12} -i\rho)^{\frac{d-4}{2} }
F_1\left(1; 1,\frac{4-d}{2}; 
2; \dfrac{M_{12} -m_2^2}{ M_{12} -M_{3} }, 
1-\dfrac{m_2^2}{M_{12} } \right)
\;\;\Bigg], \nonumber
\end{eqnarray}
provided that 
$\left|\frac{M_{12} -m_{1;2}^2}{ M_{12} -M_{3} } \right|,\;  
\left|1-\frac{m_{1;2}^2}{M_{12} }\right|<1$.

It is important to note that the results 
in (\ref{j123general1}, \ref{j123general2})
are only valid if the absolute value of 
the arguments of {the Appell $F_1$ functions} 
in these representations are less 
than $1$. If these kinematic variables do not 
satisfy this condition. 
One has to perform analytic {continuations}, 
we refer the work of~\cite{olsson} 
for Appell $F_1$.  

The results shown in
Eqs.~(\ref{j123general1}, \ref{j123general2}) 
are also new hypergeometric representations 
for scalar one-loop three-point 
functions in general space-time dimension. 
From these representations, one can perform 
analytic continuation 
of the result in the cases
of $M_{ij}=0, m_i^2, m_j^2$, $M_3=0$ 
(for $i,j,k=1,2,3$), $G_3=0$ and 
massless case, etc. This can be done 
by applying transformations for
Appell $F_1$ functions 
(seen appendix $B$).

For a example, we consider 
the case of $G_3=0$. In the case, $\beta = c/b$,  
repeating the calculation, we arrive at
\begin{eqnarray}
\label{g123J30}
J_3
= -\sum\limits_{k=1}^3 
  \left(\dfrac{\partial_k S_3}{2S_3} \right) 
  {\bf k^{-} }  J_3(d; \{p_i^2\}, \{m_i^2\} ).
\end{eqnarray}
The operator ${\bf k^{-} }$ is defined in 
such a way that 
it will reduce the three-point integrals 
to two-point integrals by shrinking
an propagator in the integrand of $J_3$. 
This equation equivalents with Eq.~($46$) 
in Ref.~\cite{Devaraj:1997es}. Noting that 
we have already arrived this relation
in (\ref{j3gram}) of previous subsection.
\subsection{Tensor one-loop three-point integrals} 
Following tensor reduction method 
in Ref.~\cite{Davydychev:1991va}, tensor 
one-loop three-point integrals with 
rank $M$ can be presented in terms 
of scalar ones with the shifted space-time 
dimension: 
\begin{eqnarray}
&& \hspace{-1cm}
J^{(3)}_{\mu_1\mu_2\cdots\mu_M} (d;\{\nu_1,\nu_2,\nu_3\}) =
\\
&=& \int \frac{d^d k}{i\pi^{d/2}} 
\dfrac{k_{\mu_1}k_{\mu_2}\cdots k_{\mu_M}}
{[(k+q_1)^2 -m_2^2 + i\rho]^{\nu_1}
[(k+q_2)^2 -m_3^2 + i\rho]^{\nu_2}
[(k+q_3)^2 -m_1^2 + i\rho]^{\nu_3}} \nonumber\\
&=&\sum\limits_{\lambda, \kappa_1, \cdots, \kappa_3} 
\left(-\frac{1}{2}\right)^{\lambda} 
\Big\{[g]^\lambda [q_1]^{\kappa_1} 
[q_2]^{\kappa_2} [q_3]^{\kappa_3} \Big\}_
{\mu_1\mu_2\cdots\mu_M}    \\
&&\hspace{1.4cm} 
\times (\nu_1)_{\kappa_1}(\nu_2)_{\kappa_2}(\nu_3)_{\kappa_3}
\;\;
J_3\Big(d+2(M-\lambda);
\{\nu_1+\kappa_1,\nu_2+\kappa_2,\nu_3+\kappa_3\} \Big).
\nonumber
\end{eqnarray}
In this formula, the condition for the indices 
$\lambda, \kappa_1, \kappa_2$ and 
$\kappa_3$ is $2\lambda +\kappa_1+\kappa_2+\kappa_3=M$. 
Moreover, these indices also follow the constrain 
$0 \leq \kappa_1, \kappa_2, \kappa_3 \leq M$ 
and $0 \leq \lambda \leq [M/2]$ 
(integer of $M/2$). The notation 
$(a)_\kappa = \Gamma(a+\kappa)/\Gamma(a)$ 
is the Pochhammer symbol. The structure of 
tensor $\{[g]^\lambda [q_1]^{\kappa_1} [q_2]^{\kappa_2} 
[q_3]^{\kappa_3} \}_{\mu_1\mu_2\cdots \mu_M} $ 
is symmetric with respect to
$\mu_1, \mu_2,\cdots, \mu_M$. 
This tensor is constructed from $\lambda$ 
of metric $g_{\mu\nu}$, 
$\kappa_1$ of momentum $q_1$, $\cdots$,  
$\kappa_3$ of momentum $q_3$.
The $J_3(d+2(M-\lambda);\{\nu_1+\kappa_1,
\nu_2+\kappa_2,\nu_3+\kappa_3\})$ are scalar 
one-loop three-point functions with the shifted 
space-time dimension 
$d+ 2(M-\lambda)$, 
raising powers of propagators 
$\{\nu_i+\kappa_i\}$ for $i=1,2,3$. 

For examples, we first take the simplest case $M=1$. 
In this case, one has $\lambda =0$.
Subsequently, we get 
\begin{eqnarray}
J^{(3)}_{\mu} (d;\{\nu_1,\nu_2,\nu_3\}) 
&=& \sum\limits_{k=1}^{3} \nu_k\; q_{k\mu} \; 
J_3(d+2;\{\nu_1 +\delta_{1k}, 
\nu_2+ \delta_{2k},\nu_3+ \delta_{3k}\}), 
\end{eqnarray}
with $\delta_{jk} $ is the Kronecker symbol. 
We next consider $M=2$. One has 
\begin{eqnarray}
J^{(3)}_{\mu_1\mu_2} (d; \{\nu_1,\nu_2,\nu_3\})
&=& -\frac{1}{2} g_{\mu_1\mu_2}\; 
J_3 (d+2;\{\nu_1, \nu_2, \nu_3\}) 
\\
&&
\hspace{-2cm}
+ \sum\limits_{k=1}^{3} q_{k\mu_1} \; q_{k\mu_2} \; (\nu_k)_2 
~
J_3 (d+4; \{\nu_1 +2\delta_{1k}, 
\nu_2 +2\delta_{2k}, \nu_3 +2\delta_{3k}\})  
\nonumber\\
&&\hspace{-2cm}
+ \sum\limits_{k=1}^{3}\sum\limits_{k'>k}^{3}
(q_{k\mu_1} q_{k'\mu_2} + q_{k\mu_2}
q_{k'\mu_1}) \; (\nu_k)_1 \;(\nu_{k'})_1 
\nonumber\\
&&\hspace{-2cm}\times
J_3 (d+4;\{\nu_1 +\delta_{1k}+\delta_{1k'}, 
\nu_2 +\delta_{2k}+\delta_{2k'}, 
\nu_3 +\delta_{3k}+\delta_{3k'}\} ). \nonumber
\end{eqnarray}
In the next step, the scalar integrals 
$J_3\Big(d+2(M-\lambda); 
\{\nu'_1,\nu'_2, \nu'_3\} \Big)$ will be 
reduced to subset of master integrals by 
using integration-by-part method 
(IBP)~\cite{IBP}. By applying the 
operator $\frac{\partial}{\partial k} \cdot k$ 
to the integrand of $J_3\left(d; \{\nu_1, \nu_2, \nu_3 \} \right)$ 
and choosing $k$ to be the momentum of three 
internal lines ($k\equiv\{k+q_1, k+q_2, k+q_3\}$). 
One arrives at the following system equations:
\begin{eqnarray}
\label{j30}
\begin{cases}
2 \nu_1 m_1^2 \mathbf{1^+} - \nu_2 Y_{12} 
\mathbf{2^+} - \nu_3 Y_{13} \mathbf{3^+} 
= (d- 2 \nu_1 -\nu_2-\nu_3) \mathbf{1} -\nu_2 
\mathbf{1^-}\mathbf{2^+} 
-\nu_3 \mathbf{1^-}\mathbf{3^+},      \\
- \nu_1 Y_{12} \mathbf{1^+} + 2 \nu_2 m_2^2 
\mathbf{2^+} - \nu_3 Y_{23} \mathbf{3^+} 
= (d-  \nu_1 - 2 \nu_2-\nu_3) \mathbf{1} - \nu_1 
\mathbf{1^+}\mathbf{2^-} 
-\nu_3 \mathbf{2^-}\mathbf{3^+},      \\
- \nu_1 Y_{13} \mathbf{1^+} - \nu_2 Y_{23} 
\mathbf{2^+} +2 \nu_3 m_3^2 \mathbf{3^+}  
= (d- \nu_1 -\nu_2-2 \nu_3) \mathbf{1} -\nu_1 
\mathbf{1^+}\mathbf{3^-} 
-\nu_2 \mathbf{2^+}\mathbf{3^-}.
\end{cases}
\end{eqnarray}
Where we used the following notation:
\begin{eqnarray}
\begin{cases}
\mathbf{1}     = J_3 (d; \{\nu_1,\nu_2,\nu_3\}), \\
\mathbf{1^\pm }= J_3 (d; \{\nu_1\pm 1,\nu_2,\nu_3\}), \\
\mathbf{2^\pm }= J_3 (d; \{\nu_1,\nu_2\pm 1,\nu_3\}), \\
\mathbf{3^\pm }= J_3 (d; \{\nu_1,\nu_2,\nu_3\pm 1\}). 
\end{cases}
\end{eqnarray}
In order to solve system equations (\ref{j30}), 
one first considers the following matrix:
\begin{eqnarray}
Y_{3} =
\begin{bmatrix}
\nu_1 Y_{11} &&& -\nu_2 Y_{12}  &&& -\nu_3 Y_{13}   \\
-\nu_1 Y_{12} &&& \nu_2 Y_{22}  &&& -\nu_3 Y_{23}   \\
-\nu_1 Y_{13} &&& -\nu_2 Y_{23}  &&& \nu_3 Y_{33}   \\
\end{bmatrix}
\text{with}\; 
Y_{ij}&=&-(q_i-q_j)^2+m_i^2+m_j^2. 
\end{eqnarray}
If det$(Y_{3}) \neq 0$, one then can present
$J_3 (d; \nu_{123}+1)$ in term of $J_3 (d; \nu_{123})$
with $\nu_{123}=\nu_1+\nu_2+\nu_3$. In this recurrence way
\cite{Laporta:2001dd}, 
we can arrive at the following integrals:
$J_2(d; \{\nu_1, \nu_2\})$
and $J_3(d; \{1,1,1\})$. By applying IPB once again
for 
the former integrals, we will arrive at master 
integrals such as:
$J_1(d; \{\nu\})$, $J_2(d; \{1,1\})$ 
which can be found in
\cite{Fleischer:2003rm} 
and $J_3(d;\{1,1,1\})$ in this paper.
\section{Conclusions}   
\noindent
New analytic formulas for 
one-loop three-point 
Feynman integrals in general space-time dimension
have presented in this paper. The results are expressed 
in terms of Appell $F_1$ functions, considered all cases 
of internal mass and external momentum assignments. We 
have also cross-checked the analytic results in this 
work with other paper in several special cases.
\\

\noindent
{\bf Acknowledgment:}~This research is funded by 
Vietnam National University (VNU-HCM), Ho Chi Minh City 
under grant number C$2019$-$18$-$06$???.  
\section*{Appendix $A$: Evaluating
the master integral}
We are considering master integral
\begin{eqnarray}
 \mathcal{K} = \int\limits_0^1 dx\; 
 \dfrac{\left[ p^2_ix^2
 - (p_i^2 +m_i^2-m_j^2)x +m_i^2 -i\rho 
 \right]^{\frac{d-4}{2} }}{x - x_k},
\end{eqnarray}
with $p_i^2 \neq 0$. Where $|x_k|>1$ 
or $-1 \leq x_k <0$ and it follows 
the below equation
\begin{eqnarray}
p^2_ix_k^2
 - (p_i^2 +m_i^2-m_j^2)x_k +m_i^2 -i\rho = M_3 -i\rho, 
\end{eqnarray}
for $i,j,k=1,2,3$. 

We will discuss on the method to evaluate this integral  
under the above conditions. In the case of $0<x_k<1$, 
one will perform analytic continuation the result for 
all master integrals which are expected to appear in  the 
general formula of $J_3$. This will be devoted in 
concrete in section $2$. 

In order to work out the master integral
one should write the polynomial of $x$ which appears in 
numerator of the $\mathcal{K}$'s integrand as 
follows
\begin{eqnarray}
 p^2_ix^2- (p_i^2 +m_i^2-m_j^2)x +m_i^2 -i\rho 
 = p^2_i(x-x_{ij})^2 +M_{ij} -i\rho.
\end{eqnarray}
We have introduced the kinematic variables
\begin{eqnarray}
 x_{ij} &=& \dfrac{p_i^2 +m_i^2-m_j^2 }{2p_i^2},                \\
 x_{ji} &=&1-x_{ij}=\dfrac{p_i^2 -m_i^2+m_j^2 }{2p_i^2},
 \end{eqnarray}
for $p_i^2 \neq 0$. From the conventions, 
we subsequently verify the below relations
\begin{eqnarray}
 \label{rij-rijk}
  p_i^2  x_{ij}^2 &=& m_i^2 - M_{ij},\\
  p_i^2  x_{ji}^2 &=& m_j^2 - M_{ij},\\
  p_i^2 (x_k-x_{ij})^2 &=& M_3 -M_{ij}, \\
  p_i^2 (1-x_k-x_{ji})^2 &=& M_3 -M_{ij}, 
 \end{eqnarray}
for $i,j =1,2,3$. 

Using Mellin-Barnes relation~\cite{MB}
we then decompose the integrand as
\begin{eqnarray}
\label{MB1}
\mathcal{K} = 
\frac{1}{2\pi i} \int\limits_{-i\infty}^{i\infty} ds\;
\dfrac{\Gamma(-s) \Gamma\left(\frac{4-d}{2} +s\right)}
{\Gamma\left(\frac{4-d}{2}\right)}
\left( \dfrac{1}{M_{ij} -i\rho }\right)^{\frac{4-d}{2} }
\int\limits_0^1  
\dfrac{dx}{x-x_{k}}
\left[\dfrac{p_i^2(x-x_{ij})^2}{M_{ij} -i\rho} \right]^s.
\end{eqnarray}
With the help of this relation, the Feynman 
parameter integral will be casted 
into the simpler form. It will be calculated 
in terms of Gauss hypergeometric 
functions. In particular, we have  
\begin{eqnarray}
\mathcal{L} = \int\limits_0^1\dfrac{dx}{x-x_{k}} 
\left[\dfrac{p_i^2(x-x_{ij})^2}{M_{ij} -i\rho} \right]^s 
= \left\{ \int\limits_0^{x_{ij}} dx 
+   \int\limits_{x_{ij}}^1 dx  \right\}
\dfrac{1}{x-x_{k}}
\left[\dfrac{p_i^2(x-x_{ij})^2}{M_{ij} -i\rho} \right]^s.
\end{eqnarray}
One makes a shift $x = x_{ij}\; x'$ 
(and $x = 1- x_{ji}\; x'$) for the first integral 
(second integral) respectively. The result reads
\begin{eqnarray}
\mathcal{L} 
&=& \left(- \dfrac{x_{ij}}{x_k} \right) 
\left(\dfrac{p_i^2 x_{ij}^2}{M_{ij} -i\rho} \right)^s
\int\limits_0^1 dx \; 
\dfrac{(1-x)^{2s}}
{1-\left(  \dfrac{x_{ij}}{x_k}\right) x} 
- \Big\{x_{ij} \leftrightarrow 
x_{ji};\;  x_k \leftrightarrow 1-x_k\Big\}-\mathrm{term}.
\nonumber\\
\end{eqnarray}
We apply a further transformation 
like $\xi = (1-x)^2 \geqslant 0$. 
The Jacobian of the shift is 
$dx = -\dfrac{d\xi}{2\sqrt{\xi}}$ 
and the integration domain is now $[1,0]$. 
As a result of this shift, we arrive at 
 \begin{eqnarray}
 \mathcal{L} &=&  -\left(\dfrac{x_{ij}}{2} \right) 
 \left(\dfrac{p_i^2 x_{ij}^2}{M_{ij} -i\rho} \right)^s
 \int\limits_0^1 d\xi \; \dfrac{\xi^{s}}{\sqrt{\xi}
 \left(x_k -x_{ij} +x_{ij}\sqrt{\xi} \right) } \\
 && \nonumber\\
 && - \Big\{x_{ij} \leftrightarrow x_{ji};\;  x_k 
 \leftrightarrow 1-x_k\Big\}-\mathrm{term} \nonumber\\
 &=& -\dfrac{x_{ij}}{2(x_k -x_{ij} )}\;
 \left(\dfrac{p_i^2 x_{ij}^2}{M_{ij} -i\rho} \right)^s
 \int\limits_0^1 d\xi \; 
 \dfrac{\xi^{s-\frac{1}{2}}}{1- 
 \dfrac{x_{ij}^2 }{ (x_k - x_{ij})^2} \xi }\nonumber\\
 &&\nonumber\\
&& + \dfrac{x_{ij}^2 }{2(x_k -x_{ij} )^2}\; 
\left(\dfrac{p_i^2 x_{ij}^2}{M_{ij} -i\rho} \right)^s
 \int\limits_0^1 d\xi \; \dfrac{\xi^{s} }{1- 
 \dfrac{x_{ij}^2 }{ (x_k - x_{ij})^2} \xi }\\
 &&\nonumber\\
 && - \Big\{x_{ij} \leftrightarrow x_{ji};\; 
 x_k \leftrightarrow 1-x_k\Big\}-\mathrm{term}. \nonumber\\
 \nonumber
 \end{eqnarray}
Following Eq.~(\ref{gauss-int}) in appendix $B$,
we can present this integral in terms of
Gauss hypergeometric functions
 \begin{eqnarray}\quad 
 \mathcal{L} &=&  -\dfrac{x_{ij}}{2(x_k -x_{ij} )}
 \;\left(\dfrac{p_i^2 x_{ij}^2}{M_{ij} -i\rho} \right)^s
 \dfrac{\Gamma\left(s+\frac{1}{2} \right)}
 { \Gamma\left(s+\frac{3}{2} \right)}
 \Fh21\Fz{s + \frac{1}{2}, 1}{s+ \frac{3}{2}}
 { \dfrac{x_{ij}^2 }{ (x_k - x_{ij})^2} } \nonumber\\
 && +\dfrac{x_{ij}^2 }{2(x_k -x_{ij} )^2} 
 \; \left(\dfrac{p_i^2 x_{ij}^2}{M_{ij} -i\rho} \right)^s
 \dfrac{\Gamma\left(s+1 \right)}{ \Gamma\left(s+2 \right)}
 \Fh21\Fz{s + 1, 1}{s+ 2}{\dfrac{x_{ij}^2 }{ (x_k - x_{ij})^2} } 
 \\
 &&\nonumber\\
 &&-\Big\{x_{ij} 
 \leftrightarrow x_{ji};\;  
 x_k \leftrightarrow 1-x_k\Big\}-\mathrm{term},
 \nonumber
 \end{eqnarray}
provided that 
$\left|\frac{x_{ij}^2 }{ (x_k - x_{ij})^2}\right|<1$
for $i,j,k=1,2,3$
and $\mathcal{R}$e$\left(s+\frac{1}{2}\right)>0$.

Putting this result into Eq.~(\ref{MB1}), we are going
to evaluate the following Mellin-Barnes integral
\begin{eqnarray}
\label{MB2}
\dfrac{1}{2\pi i} \int\limits_{-i\infty}^{+i\infty}ds \;
\dfrac{\Gamma(-s)\;\Gamma(a +s) \;\Gamma(b+s)}{\Gamma(c+s) }
(-x)^s \; \Fh21\Fz{a+s,b'}{c+s}{y},  
\end{eqnarray}
with $|\mathrm{Arg}(-x)|<\pi$ and $|x|<1$ and $|y|<1$. 
Under these conditions,
one could close the integration contour to 
the right side of the imaginary axis in the
$s$-complex plane. Subsequently, we take 
into account the residua of the 
sequence poles of $\Gamma(-s)$. The result
is presented as a series of Appell $F_1$ 
functions~\cite{Slater}
\begin{eqnarray}
\label{MB3}
\dfrac{\Gamma(a)\Gamma(b)}{\Gamma(c)} \sum\limits_{m=0}^{\infty}
\dfrac{(a)_m (b)_m}{(c)_m} \dfrac{x^m}{m!} \Fh21\Fz{a+m,b'}{c+m}{y}
=\dfrac{\Gamma(a)\Gamma(b)}{\Gamma(c)}\; F_1 (a; b, b';c; x,y), 
\nonumber
\end{eqnarray}
provided that $|\mathrm{Arg}(-x)|<\pi$ 
and $|x|<1$ and $|y|<1$. Finally, the result 
for $\mathcal{K}$ reads
\begin{eqnarray}
\label{K1}
\mathcal{K} &=&  -\dfrac{x_{ij}}{(x_k -x_{ij} )}\;
\left(M_{ij} -i\rho\right)^{\frac{d-4}{2} } 
\; F_1\left(\frac{1}{2}; 1, \frac{4-d}{2}; 
\frac{3}{2}; \dfrac{x_{ij}^2 }{ (x_k - x_{ij})^2}, 
-\dfrac{p_i^2 x_{ij}^2}{M_{ij} -i\rho}  \right)\nonumber\\
&&+ \dfrac{x^2_{ij}}{2 (x_k -x_{ij} )^2}\; 
\left(M_{ij} -i\rho\right)^{\frac{d-4}{2} } 
\; F_1\left(1; 1, \frac{4-d}{2}; 2; 
\dfrac{x_{ij}^2 }{ (x_k - x_{ij})^2},
-\dfrac{p_i^2 x_{ij}^2}{M_{ij}-i\rho }  \right)  
\nonumber   \\
&&\nonumber \\
&&-\Big\{x_{ij} \leftrightarrow x_{ji};
\; x_k \leftrightarrow 1-x_k\Big\}-\mathrm{term}, 
\end{eqnarray}
provided that $\left|\frac{x_{ij}^2 }{ (x_k - x_{ij})^2} \right| <1$ 
and $ \left|-\frac{p_i^2 x_{ij}^2}{M_{ij}} \right| <1$ for $i,j,k=1,2,3$.
It is confirmed that 
Arg$\left( \frac{x_{ij}^2 }{ (x_k - x_{ij} )^2 }\right)<\pi$
and Arg$\left(-\frac{p_i^2 x_{ij}^2}{M_{ij}}\right)<\pi$ 
due to the $i\rho$-term. 
In other words, these kinematic variables are 
never on the negative real axis. 

Applying the transformation for Appell $F_1$ functions 
(see Eq.~(\ref{f1relation1}) in appendix $B$, one finds 
another representation for integral $\mathcal{K}$ as
\begin{eqnarray}
\label{K2}
 \mathcal{K} &=& 
 \dfrac{x_{ij} (x_{ij}-x_k)
 \left(m_i^2 \right)^{\frac{d-4}{2} }
 }{ (x_k -x_{ij} )^2 - x_{ij}^2 }
 \; F_1\left(1; 1, \frac{4-d}{2}; 
 \frac{3}{2}; \dfrac{-x_{ij}^2}{ (x_k -x_{ij} )^2 -x_{ij}^2 }, 
\frac{p_i^2 x_{ij}^2 }{p_i^2 x_{ij}^2+M_{ij} -i\rho}  \right)
\nonumber\\
 && + \dfrac{x^2_{ij} \left(m^2_{i} \right)^{\frac{d-4}{2}} 
 }{2 \left[ (x_k -x_{ij} )^2 -x_{ij}^2\right] }\; 
 F_1\left(1; 1, \frac{4-d}{2}; 2;
 \dfrac{-x_{ij}^2}{ (x_k -x_{ij} )^2 -x_{ij}^2 }, 
\frac{p_i^2 x_{ij}^2 }{p_i^2 x_{ij}^2   +M_{ij}-i\rho }  
\right)         \nonumber \\
 &&\nonumber\\
 &&-\Big\{x_{ij} \leftrightarrow x_{ji};\; 
 x_k \leftrightarrow 1-x_k\Big\}-\mathrm{term}, 
\end{eqnarray}
provided that $ \left|-\frac{x_{ij}^2} 
{ (x_k -x_{ij} )^2 -x_{ij}^2 } \right| <1$ and
$ \left| \frac{p_i^2 x_{ij}^2 }
{p_i^2 x_{ij}^2   +M_{ij} } \right| <1$ for $i,j,k=1,2,3$.
\section*{Appendix $B$: 
Generalized hypergeometric functions}
The Gauss hypergeometric series are given 
(see Eq.~($1.1.1.4$) in Ref.~\cite{Slater})
\begin{eqnarray}
\label{gauss-series}
\Fh21\Fz{a,b}{c}{z} 
= \sum\limits_{n=0}^{\infty} 
\dfrac{(a)_n (b)_n}{(c)_n} \frac{z^n}{n!},
\end{eqnarray}
provided that $|z|<1$. Here, the pochhammer symbol, 
\begin{eqnarray} 
(a)_n =\dfrac{\Gamma(a+n)}{\Gamma(a)},
\end{eqnarray} 
is taken into account. 
The integral representation for Gauss hypergeometric 
functions is (see Eq.~(1.6.6) in Ref.~\cite{Slater})
\begin{eqnarray}
\label{gauss-int}
\Fh21\Fz{a,b}{c}{z} 
=\dfrac{\Gamma(c)}{\Gamma(b)\Gamma(c-b)} \int\limits_0^1 du
\; u^{b-1} (1-u)^{c-b-1} (1-zu)^{-a},
\end{eqnarray}
provided that  $|z|<1$ and Re$(c)>$Re$(b)>0$.

The series of Appell $F_1$ functions are  given 
(see Eq.~(8.13) in Ref.~\cite{Slater})
\begin{eqnarray}
\label{appell-series}
F_1(a; b, b'; c; x, y) 
= \sum\limits_{m=0}^{\infty}\sum\limits_{n=0}^{\infty}
  \dfrac{(a)_{m+n} (b)_m (b')_n}{(c)_{m+n}\; m! n!} x^m y^n,
\end{eqnarray}
provided that $|x|<1$ and $|y|<1$.
The single integral representation for $F_1$ is 
(see Eq.~(8.25) in Ref.~\cite{Slater})
\begin{eqnarray}
\label{appell-int}
F_1(a; b, b'; c; x, y) 
=  \dfrac{\Gamma(c)}{\Gamma(c-a)\Gamma(a)} \int\limits_0^1 du\;
   u^{a-1} (1-u)^{c-a-1} (1-xu)^{-b}(1-yu)^{-b'}, 
\end{eqnarray}
provided that Re$(c)$ $>$ Re$(a)>0$ and $|x|<1$, $|y|<1$.
\subsection*{Transformations for Gauss    
$_2F_1$ hypergeometric functions}         
Basic linear transformation formulas for 
Gauss $_2F_1$ hypergeometric functions 
collected from Ref.~\cite{Slater} 
are listed as follows
\begin{eqnarray}
\Fh21\Fz{a,b}{c}{z} &=& \Fh21\Fz{b,a}{c}{z} 
\label{tran2F1a}   \\
&=& (1-z)^{c-a-b} \Fh21\Fz{c-a,c-b}{c}{z} 
\label{tran2F1b} \\
&=& (1-z)^{-a} \Fh21\Fz{ a,c-b}{c}{\frac{z}{z-1} } 
\label{tran2F1c} \\
&=& (1-z)^{-b} \Fh21\Fz{ b,c-a}{c}{\frac{z}{z-1} } 
\label{tran2F1d}\\
&=& \frac{\Gamma(c) \Gamma(c-a-b)}{\Gamma(c-a) \Gamma(c-b)} 
\Fh21\Fz{a,b}{a+b-c+1}{1-z} \nonumber \\
&+& (1-z)^{c-a-b} \frac{\Gamma(c) \Gamma(a+b-c)}{\Gamma(a) \Gamma(b)} 
\Fh21\Fz{c-a,c-b}{c-a-b+1}{1-z}\\
&=& \frac{\Gamma(c) \Gamma(b-a)}{\Gamma(b) \Gamma(c-a)} (-z)^{-a} 
\Fh21\Fz{a,1-c+a}{1-b+a}{\frac{1}{z} } \nonumber \\
&+& \frac{\Gamma(c) \Gamma(a-b)}{\Gamma(a) \Gamma(c-b)} (-z)^{-b} 
\Fh21\Fz{b,1-c+b}{1-a+b}{\frac{1}{z} }\label{z-}.
\end{eqnarray}
\subsection*{Transformations for Appell 
$F_1$ hypergeometric functions}
We collect all transformations for Appell 
$F_1$ functions from 
Refs.~\cite{Slater}. The first 
relation for $F_1$ is mentioned, 
\begin{eqnarray}
\label{f1relation1}
F_1\Big(a;b,b';c;x,y\Big)=(1-x)^{-b}(1-y)^{-b'}
F_1\Big(c-a;b,b';c;\frac x{x-1},\frac y{y-1}\Big).
\end{eqnarray}
If $b'=0$, we arrive at the well-known 
Pfaff--Kummer transformation for the $_2F_1$. 
In detail, one has
\begin{eqnarray}
F_1\Big(a;b,b';c;x,y\Big)=(1-x)^{-a}
F_1\!\left(a;-b-b'+c,b';c;\frac x{x-1},\frac{y-x}{1-x}\right).
\end{eqnarray}
Furthermore, if $c=b+b'$, one then obtains
\begin{eqnarray}
F_1\Big(a;b,b';b+b';x,y\big)&=&(1-x)^{-a}
\Fh21\Fz{a,b'}{b+b'}{\frac{y-x}{1-x}}
\\
&=&
(1-y)^{-a}\Fh21\Fz{a,b}{b+b'}{\frac{x-y}{1-y}}.
\end{eqnarray}
Similarly,
\begin{eqnarray}
 F_1\left(a;b,b';c;x,y\right)&=&(1-y)^{-a}
 F_1\!\left(a;b,c-b-b';c;\frac{x-y}{1-y},\frac y{y-1}\right),
 \\
 F_1\left(a;b,b';c;x,y\right) &=&
 {\mbox{\small $(1-x)^{c-a-b}(1-y)^{-b'}$}}
 F_1\!\left(c-a;c-b-b',b';c;x,\frac{x-y}{1-y}\right), 
\\
 F_1\left(a;b,b';c;x,y\right) &=&
 {\mbox{\small $(1-x)^{-b}(1-y)^{c-a-b'}$}}
 F_1\!\left(c-a;b,c-b-b';c;\frac{y-x}{1-x},y\right).
 \end{eqnarray}

\end{document}